\documentclass{aa}  

\usepackage{natbib}
\usepackage{graphicx}
\usepackage{amsmath}
\usepackage{txfonts}
\usepackage{hyperref}
\usepackage{xspace}

\def\mobs{M_*^{\mathrm{(obs)}}}
\def\mobsi{M_{*,i}^{\mathrm{(obs)}}}
\def\mstar{M_*}
\def\mstari{M_{*, i}}

\def\mhalo{M_{h}}
\def\mhaloi{M_{h, i}}

\def\reff{R_{\mathrm{e}}}
\def\reffi{R_{\mathrm{e},i}}
\def\robs{R_{\mathrm{e}}^{\mathrm{(obs)}}}
\def\robsi{R_{\mathrm{e},i}^{\mathrm{(obs)}}}

\def\rsky{R_{\mathrm{sky}}}

\def\nser{n}
\def\mhukpc{M_{*,100\mathrm{kpc}}}
\def\mtenkpc{M_{*,10\mathrm{kpc}}}

\def\hyperp{\boldsymbol{\eta}}

\def\data{\mathbf{d}}
\def\datai{\mathbf{d}_i}

\def\Sref#1{Section~\ref{#1}\xspace}
\def\Fref#1{Figure~\ref{#1}\xspace}
\def\Tref#1{Table~\ref{#1}\xspace}
\def\Eref#1{Equation~\ref{#1}\xspace}

\def\Nlens{10,403}

\def\pr{{\rm P}}

\begin{document}

   \title{An HSC view of the CMASS galaxy sample. Halo mass as a function of stellar mass, size and S\'{e}rsic index.}
   \titlerunning{An HSC view of the CMASS sample.}
   \authorrunning{Sonnenfeld et al.}


   \author{Alessandro Sonnenfeld\inst{1,2}\thanks{Marie Sk\l{}odowska-Curie Fellow}\and
           Wenting Wang\inst{2}
	   \and
	   Neta Bahcall\inst{3}
          }

   \institute{Leiden Observatory, Leiden University, Niels Bohrweg 2, 2333 CA Leiden, the Netherlands\\
              \email{sonnenfeld@strw.leidenuniv.nl} \and
   Kavli IPMU (WPI), UTIAS, The University of Tokyo, Kashiwa, Chiba 277-8583, Japan\\
   \and
   Department of Astrophysical Sciences, Princeton University, 4 Ivy Lane, Princeton, NJ 08544, USA}

   \date{}

 
  \abstract
   {}
   {We wish to determine the distribution of dark matter halo masses as a function of the stellar mass and the stellar mass profile, for massive galaxies in the BOSS CMASS sample.} 
   {We use $grizy$ photometry from HSC to obtain S\'{e}rsic fits and stellar masses of CMASS galaxies for which HSC weak lensing data is available, visually selected to have spheroidal morphology. 
We apply a cut in stellar mass, $\log{M_*/M_\odot} > 11.0$, selecting $\sim10,000$ objects.
Using a Bayesian hierarchical inference method, we first investigate the distribution of S\'{e}rsic index and size as a function of stellar mass. Then, making use of shear measurements from HSC, we measure the distribution of halo mass as a function of stellar mass, size and S\'{e}rsic index.
}
   {Our data reveals a steep stellar mass-size relation $R_e \propto M_*^{\beta_R}$, with $\beta_R$ larger than unity, and a positive correlation between S\'{e}rsic index and stellar mass: $\nser \propto \mstar^{0.46}$.
Halo mass scales approximately with the $1.7$ power of the stellar mass. We do not find evidence for an additional dependence of halo mass on size or S\'{e}rsic index at fixed stellar mass.
}
   {Our results disfavour galaxy evolution models that predict significant differences in the size growth efficiency of galaxies living in low and high mass halos.}


   \maketitle
%

\section{Introduction}\label{sect:intro}

The size evolution of massive quiescent galaxies is one of the open problems in cosmology. 
On the one hand, observations show that, at fixed stellar mass, the average size of quiescent galaxies has increased by a factor of a few between $z\approx2$ and the present \citep[e.g.][]{Dad++05, Tru++06, van++08}.
On the other hand, theoretical models have struggled to reproduce this trend \citep[see e.g.][]{Hop++10}. 
Qualitatively, the size evolution of massive quiescent galaxies is understood to be the result of mergers. Minor mergers, in particular, are known to be an efficient mechanism to increase the size of a galaxy \citep{Naa++09}. However, it is not clear whether the observed merger rates are sufficient to explain the size evolution signal, especially in the redshift range $1 < z < 2$ \citep{New++12}.
Moreover, it is very difficult to preserve the tightness of the stellar mass-size relation observed at $z\sim0$ with evolutionary models based exclusively on dissipationless (dry) mergers \citep{Nip++12}.

Clues on the physical mechanisms that are relevant for the size evolution of massive galaxies can be obtained by studying the dependence of size on environment \citep[see][]{Sha++14}. Although the correlation between environment and galaxy size at fixed stellar mass has been investigated by a relatively large number of studies, no clear consensus has been reached, with some works showing evidence for a positive correlation \citet{Coo++12, Pap++12, Lan++13, Yoo++17, Hua++18}, and others finding results consistent with no dependence \citet{Hue++13b, New++14, All++15, Dam++15, Sar++17}.
The signal, if present, is in any case small: reported differences between the average size of galaxies found in clusters and in less massive associations are on the order of 20\%.

A possible way forward is looking at correlations between the observed properties of galaxies and those of their host dark matter halos.
The evolution of the mass of a dark matter halo traces directly its accretion and merger history. This, in turn, should relate directly to the size evolution of its central galaxy, since mergers are thought to be the main driver of the growth in size, at least at the massive end of the galaxy distribution.

Halo masses can be measured using weak gravitational lensing. However, the low strength of the weak lensing signal around typical massive galaxies requires the statistical combination of measurements over hundreds or thousands of lenses, making it challenging to obtain an accurate description of the distribution of halo mass as a function of galaxy properties other than stellar mass.

Recently, \citet{Cha++17} reported the detection of a positive correlation between halo mass and size at fixed stellar mass, using stacked weak lensing measurements from the CFHTLenS survey \citep{Hey++12}.
Here we use a Bayesian hierarchical inference method to fit for the distribution of halo mass as a function of stellar mass, size and S\'{e}rsic index, using photometric and weak lensing data from the Hyper Suprime-Cam (HSC) survey \citep{Aih++18a}.

We draw the lens galaxies used in our study from the CMASS sample of the Baryon Oscillation Spectroscopic Survey \citep[BOSS][]{SWE09, Daw++13}.
The CMASS sample has been used for a variety of studies in galaxy evolution and cosmology \citep[e.g.]{And++14, Beu++14, Mor++15, Mon++16, Lea++17, Tin++17, Fav++18}.
Thanks to the image quality ($0.6''$ typical $i-$band seeing) and depth of HSC data, our work will allow us to obtain a more accurate description of the light distribution in CMASS galaxies, compared to previous studies.
A by-product of our analysis is then a measurement of the stellar mass-size relation of CMASS galaxies, which we generalise to a stellar mass-size-S\'{e}rsic index relation, based on unprecedented deep and sharp data.

The structure of this work is the following.
In \Sref{sect:data} we describe the data used for our study.
In \Sref{sect:masssize} we carry out an investigation on the distribution of S\'{e}rsic index and half-light radius as a function of stellar mass.
In \Sref{sect:halos} we perform a weak lensing analysis to infer the distribution of dark matter halo mass as a function of various galaxy properties.
We discuss our results in \Sref{sect:discuss} and conclude in \Sref{sect:concl}.
We assume a flat $\Lambda$CDM cosmology with $\Omega_M=0.3$ and $H_0=70\,\rm{km}\,\rm{s}^{-1}\,\rm{Mpc}^{-1}$.


\section{Data}\label{sect:data}

\subsection{The CMASS sample}

The set of galaxies subject of our study is drawn from the CMASS sample of BOSS, which is part of the Sloan Digital Sky Survey III \citep[SDSS-III][]{Eis++11}.
The CMASS ('constant mass') sample has been constructed with the primary goal of selecting luminous red galaxies in the redshift range $0.43 < z < 0.7$ for the measurement of the baryon acoustic oscillation signal.
CMASS galaxies are selected by applying a series of cuts on colours and magnitudes, as measured from SDSS photometry. We refer to \citet{Rei++16} for details on the selection criteria of the CMASS sample.

For each galaxy in the CMASS sample, we have a spectroscopic redshift measured with the BOSS spectrograph.

\subsection{HSC photometry}

We select CMASS galaxies that lie in the region covered by the first-year shear catalogue of the HSC survey \citep{Man++18}. This corresponds to 136.9~deg$^2$ imaged in $grizy$ bands at the full survey depth \citep[$i\sim26.4$ mag $5\sigma$ limit for a point source,][]{Aih++18b}, resulting in a sample size of approximately 18,000 CMASS galaxies.
For each galaxy in the sample, we obtain small cutouts of coadded images in each band from the S17a internal data release. Data reduction, including the generation of coadded images, model point spread function (PSF) kernels and sky subtraction is performed using {\sc hscPipe} \citep{Bos++18}, a version of the Large Synoptic Survey Telescope software stack \citep{Ive++08, Axe++10, Jur++15}. We refer to \citet{Bos++18} for details on the data reduction process.

\begin{figure*}
\includegraphics[width=\textwidth]{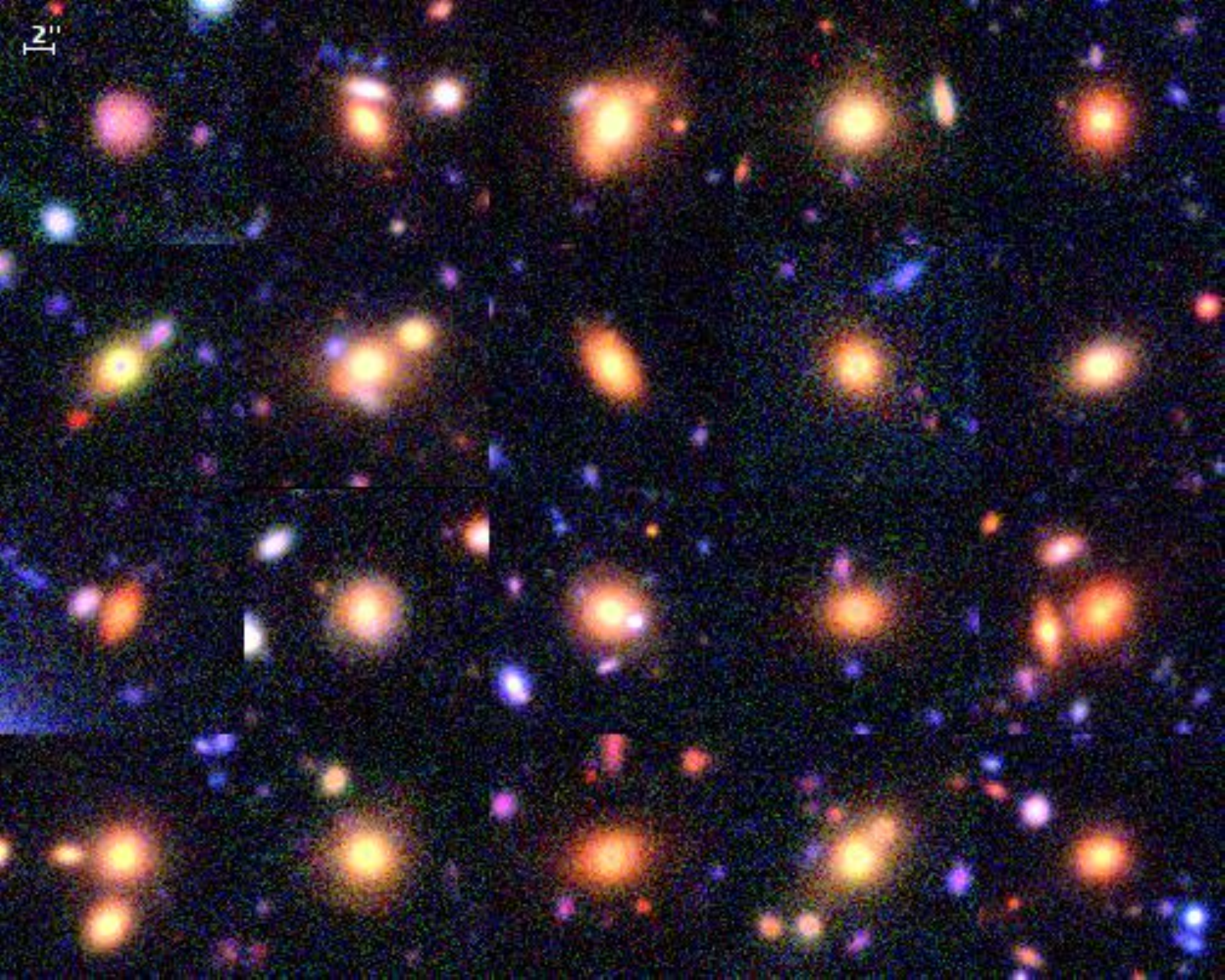}
\caption{Colour-composite images in HSC-$irg$ bands of a set of 20 galaxies drawn randomly from the sample used for our study. We used the algorithm of \citet{Mar++16} to create the images.}
\label{fig:collage}
\end{figure*}

In \Fref{fig:collage}, we show colour-composite images of a set of 20 galaxies drawn randomly from the final sample used for our study (the criteria used to define the final sample will be listed in the rest of this Section).

\subsection{S\'{e}rsic profile fits}\label{ssec:sersicprofile}

We fit an elliptical S\'{e}rsic surface brightness profile \citep{Ser68} to the photometric data in each band:
\begin{equation}
I(x, y) = I_0 \exp{\left\{-b(n)\left(\frac{R}{R_e}\right)^{1/n}\right\}},
\end{equation}
where $x$ and $y$ are Cartesian coordinates aligned with the major and minor axis of the elliptical isophotes and origin in the centre, $R$ is the circularised radius,
\begin{equation}
R^2 \equiv qx^2 + \frac{y^2}{q},
\end{equation} 
$n$ is the S\'{e}rsic index, and $b(n)$ is a numerical constant that ensures that the light enclosed within the isophote with $R=R_e$ is half of the total light \citep[see][]{C+B99}.

For each galaxy, we fix the parameters of the S\'{e}rsic profile to be the same in all five bands, with the exception of the amplitude $I_0$. In other words, we assume galaxies to have spatially uniform colours.
We run a Markov Chain Monte Carlo (MCMC) to sample the parameter space defined by the galaxy centroid, half-light radius, axis ratio, position angle, S\'{e}rsic index, and $g-i$, $r-i$, $i-z$, $y-i$ colours. 
For each set of values of these structural parameters, we find the $i-$band amplitude that minimises the following $\chi^2$,
\begin{equation}
\chi^2 = \sum_\lambda \sum_j \left(\frac{I_{\lambda,j}^{\mathrm{(mod)}} - I_{\lambda,j}^{\mathrm{(obs)}}}{\sigma_j^2}\right)^2,
\end{equation}
where $I_{\lambda,j}^{\mathrm{(mod)}}$ is the PSF-convolved S\'{e}rsic model in band $\lambda$ evaluated at pixel $j$, $I_{\lambda,j}$ is the observed surface brightness at the corresponding pixel and $\sigma_{\lambda,j}$ is the observational uncertainty.
Since the model surface brightness is linear in the $i-$band amplitude, the above $\chi^2$ can be minimised analytically.

A circular region of 50 pixel (8.4'') radius centred on each galaxy is used for the fit.
This corresponds to a physical radius of 54~kpc at $z=0.55$, the median redshift of our sample.
The choice of this aperture is a compromise between the need for capturing as much light as possible from each galaxy and at the same time avoiding to select too big of a region, in order not to be affected by systematics in the sky subtraction.
We verified that our results do not change if we use a fitting region a factor of two larger or smaller. 

The model parameters are initialised to the values of the {\sc cmodel\_dev} fit produced by {\sc hscPipe}, with $n=4$.
We run {\sc SExtractor} \citep{B+A96}, with a $2\sigma$ detection threshold, on the $i$ and $g$ band image to mask objects not associated with the main galaxy. 
We run a preliminary MCMC on the centroid and colour parameters while keeping the S\'{e}rsic index, position angle, axis ratio and half-light radius fixed, run {\sc SExtractor} once again on the model-subtracted image, then run another MCMC over the full set of parameters.
We use the package {\sc emcee} \citep{For++13} to perform the MCMC.
Typical statistical uncertainties are 2\% on $R_e$ and $n$, 1\% on the axis ratio $q$, $\sim1$~deg on the position angle and $0.01$~mag on the magnitudes.

We wish to carry out our measurements on passive galaxies only, since star-forming galaxies have been shown to lie on a different stellar-to-halo mass relation \citep{Man++16} and could therefore bias our inference.
We then visually inspect each CMASS galaxy and remove from the sample any object that shows clear signs of star formation activity or disk-like morphology. We also discard systems on which it is difficult to carry out photometric measurements in an automated way, such as blended objects, images with significant contamination from a nearby bright star, or strong gravitational lenses.
After this visual inspection step, we are left with $\sim13,000$ elliptical galaxies with clean photometry.

\subsection{Stellar masses}\label{ssec:ssp}

We estimate stellar masses by fitting synthetic stellar population models to the observed $grizy$ magnitudes, following the procedure described by \citet{Aug++09}, with minor modifications.
We take simple stellar population (SSPs, or instantaneous-burst) models, produced by \citet{B+C03} using semi-empirical stellar spectra from the BaSeL 3.1 library \citep{Wes++02} and the Padova 1994 evolutionary tracks, assuming a Chabrier initial mass function \citep[IMF,][]{Cha03}.
We then use the {\sc bc03} code \citep{B+C03} to create composite stellar population (CSP) model spectra, assuming an exponentially decaying star formation history, over a four-dimensional grid of age (i.e. time since the initial burst), star formation rate decay time $\tau$, metallicity $Z$ and dust attenuation $\tau_V$.
We use these synthetic spectra to evaluate the model broadband flux in each HSC filter at each point of the grid, adding redshift as a fifth dimension.
Finally, we fit for stellar mass, age, $\tau$, metallicity and dust attenuation by running an MCMC: for each set of values of the model parameters, we obtain the predicted $grizy$ magnitudes by interpolating over the grid, and compare them to the observed values.

\begin{figure}
\centering
\includegraphics[width=\columnwidth]{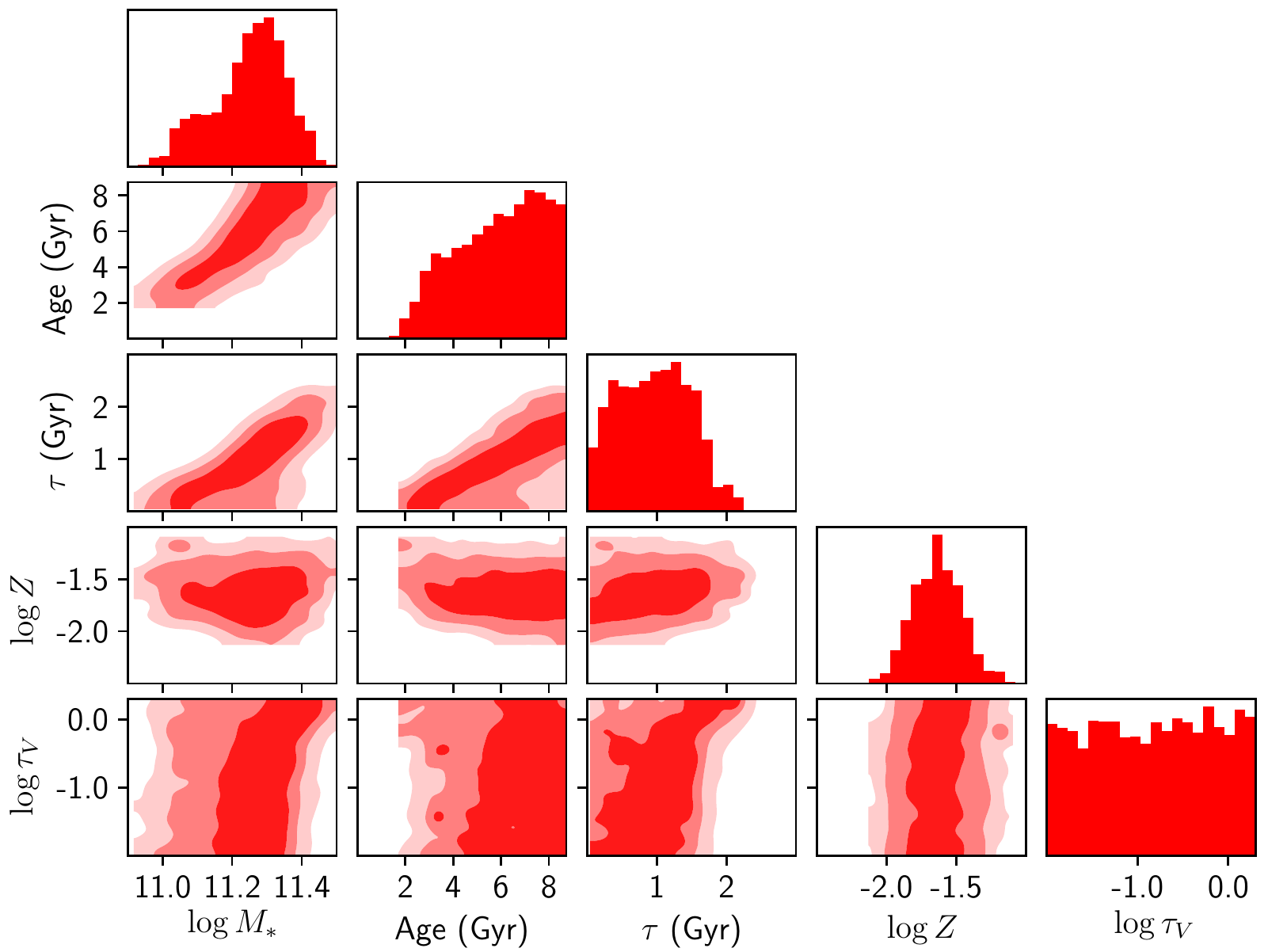}
 \caption{Posterior probability distribution of the stellar population parameters of an example galaxy. The three different contour levels mark the 68\%, 95\% and 99.7\% enclosed probability regions.}
 \label{fig:stellarpop}
\end{figure}

Thanks to the depth of HSC data, the typical observational uncertainties on the fluxes are $\sim0.01$~mag. This is a much smaller value than the systematic uncertainties in the stellar population synthesis models: in most cases, it is not possible to fit the observed magnitudes down to the noise. 
Ignoring this source of error can lead to underestimating the uncertainty on the stellar mass: a bad fit can return artificially small uncertainties, as is the case when combining measurements that are inconsistent with each other.
We then add in quadrature a $0.05$~mag uncertainty to the measurement in each band, as this is the typical scatter between the best-fit model spectral energy distribution (SED) and the data. 
We assume flat priors for all stellar population parameters, over the range covered by the \citet{B+C03} models, with the exception of metallicity, for which we adopt a truncated Gaussian prior, with mean and width that scale with stellar mass, as indicated by the observational study of \citet{Gal++05}.
In \Fref{fig:stellarpop}, we show the posterior probability distribution of the stellar population parameters for an example galaxy.

\subsection{Weak lensing data and cuts}\label{ssec:wldata}

We take weak lensing measurements from the first-year shape catalogue of the HSC survey \citep{Man++18}.
The source number density of this catalogue is $24.6$~arcmin$^{-2}$ (unweighted).
Shapes are measured on coadded $i-$band images using the re-Gaussianization PSF correction method of \citet{H+S03}. We refer to \citet{Man++18} for details about the shape measurement process and the properties of the shape catalogue.

We use photometric redshifts (photo-zs) obtained with the photo-z code {\sc Mizuki} \citep{Tan15} on the data release 1 of the HSC survey \citep{Tan++18}.

The main weak lensing analysis is carried out using the Bayesian hierarchical inference method of \citet[][SL18 from now on]{S+L18}.
This method is based on the {\em isolated lens assumption}: each lens galaxy in the sample is assumed to be at the centre of its dark matter halo, with no other mass component affecting the shape of the background sources around it.
This assumption breaks down for satellite galaxies. It is therefore important to select a sample of lenses with the lowest possible satellite fraction.
For this purpose, we first apply a cut in stellar mass, selecting only galaxies with observed stellar mass $\mobs$, defined as the median of the posterior probability distribution in $M_*$, larger than $10^{11}M_\odot$. 90\% of the CMASS galaxies left in our sample after the visual inspection step satisfy this requirement.
We then match our sample with the HSC cluster catalogue of \citep{Ogu++18}, and remove objects with a cluster membership probability larger than 50\% that are not identified as the brightest cluster galaxy.
This step removes $\sim5$\% of the objects in the sample.

Following the SL18 method, we model the weak lensing signal produced by each lens on a set of background sources within a cone of a given radius, which we take to be $300$ physical kpc at the redshift of the lens, where the isolated lens assumption is more realistic (i.e. where the so-called $1-$halo term dominates). 
One of the foundations of the SL18 method is the ability of treating the likelihood of the weak lensing data around the lenses as independent from each other, which simplifies the problem greatly.
However, in case of lenses in close proximity along the line of sight, the lensing cones of different lenses can overlap. 
In principle, one should simultaneously model the contribution of each lens on all the sources within the overlapping cones. In practice, this is computationally difficult.
We avoid this problem by eliminating overlapping pairs, using the same procedure adopted by SL18, as follows.
We rank the lenses in decreasing order of observed stellar mass, loop through them, and remove from the sample any lens the lensing cone of which overlaps with that of a more massive galaxy. The idea is that, in case of lenses in close line of sight proximity, the main lensing signal should come from the most massive galaxy.
Although in reality the problem is more complex (for instance, the strength of the lensing signal depends not only on the lens mass but also on its redshift), the fraction of objects removed with this procedure is only 10\%, and we do not expect this step to introduce any significant bias to our weak lensing analysis (as already shown by SL18 on mock observations).
We then reach our final sample size of $\Nlens$ CMASS galaxies.

\begin{figure}
\includegraphics[width=\columnwidth]{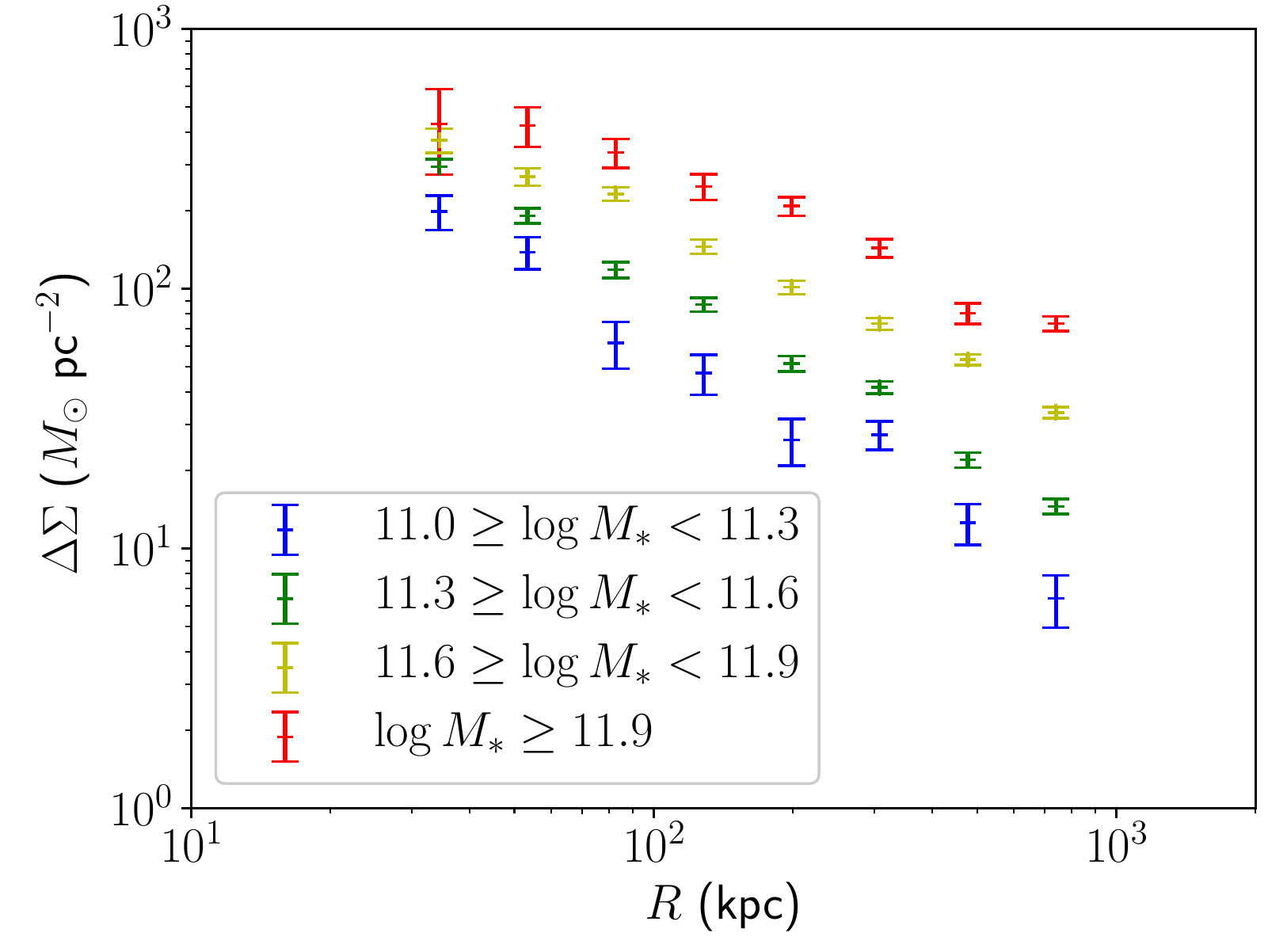}
\caption{Excess surface mass density profile in different bins of observed stellar mass, obtained by stacking the weak lensing signal in circular annuli.
}
\label{fig:stacked}
\end{figure}

To get a sense of the quality of the weak lensing data available for our study, we carry out a stacked analysis. We make four bins in stellar mass and measure the stacked excess surface mass density $\Delta\Sigma$ in different radial bins, using the HSC weak lensing pipeline.
We plot the $\Delta\Sigma$ profiles in \Fref{fig:stacked}. Differences in the lensing signal in different bins are detected with high confidence: this suggests that the data should allow us to determine not only an average halo mass of the sample, but also how halo mass scales with stellar mass and, possibly, with size.
In any case, \Fref{fig:stacked} is only shown for illustration purposes: all our weak lensing analysis is carried out by forward-modelling the weak lensing signal of individual halos.

\subsection{The final sample}

\begin{figure}
\centering
\includegraphics[width=\columnwidth]{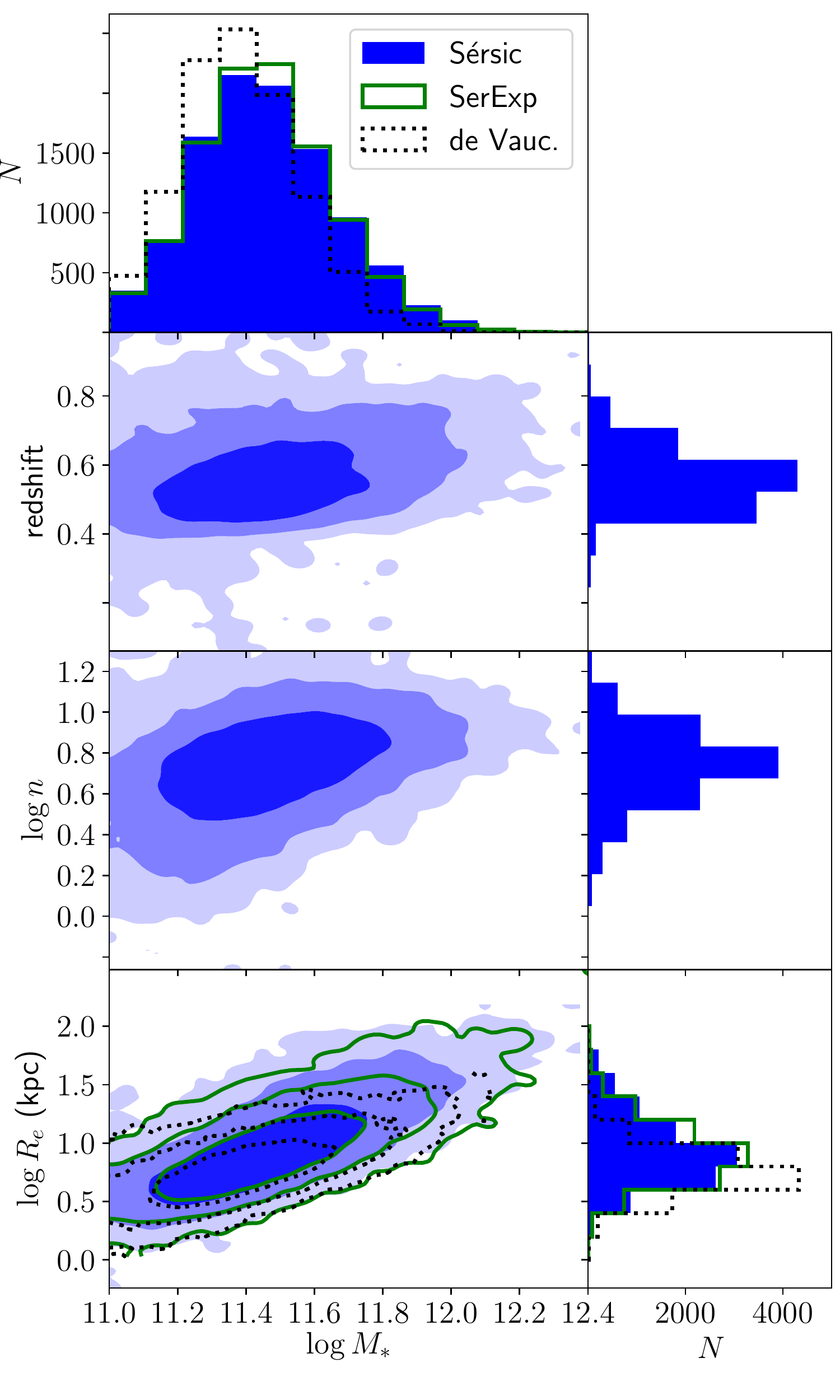}
 \caption{Distribution in half-light radius, S\'{e}rsic index and redshift as a function of stellar mass of the sample of $\Nlens$ CMASS galaxies subject of our study. Contour levels mark the 68\%, 95\% and 99.7\% enclosed probability regions. Blue contours and histograms refer to the fiducial model, consisting of a single S\'{e}rsic surface brightness profile for each galaxy. The distribution in stellar mass and size obtained with the SerExp model, described in subsection \ref{ssec:serexp}, is plotted in green, while that obtained with the de Vaucouleurs model is marked by solid lines.}
 \label{fig:mass_size}
\end{figure}

In \Fref{fig:mass_size}, we plot the distribution in observed stellar mass, half-light radius, S\'{e}rsic index and redshift of the final sample.
We can see a very clear correlation between mass and size, and also a trend of increasing S\'{e}rsic index with stellar mass. In the next Section we will quantify the strength of these correlations.

\begin{figure}
\centering
\includegraphics[width=\columnwidth]{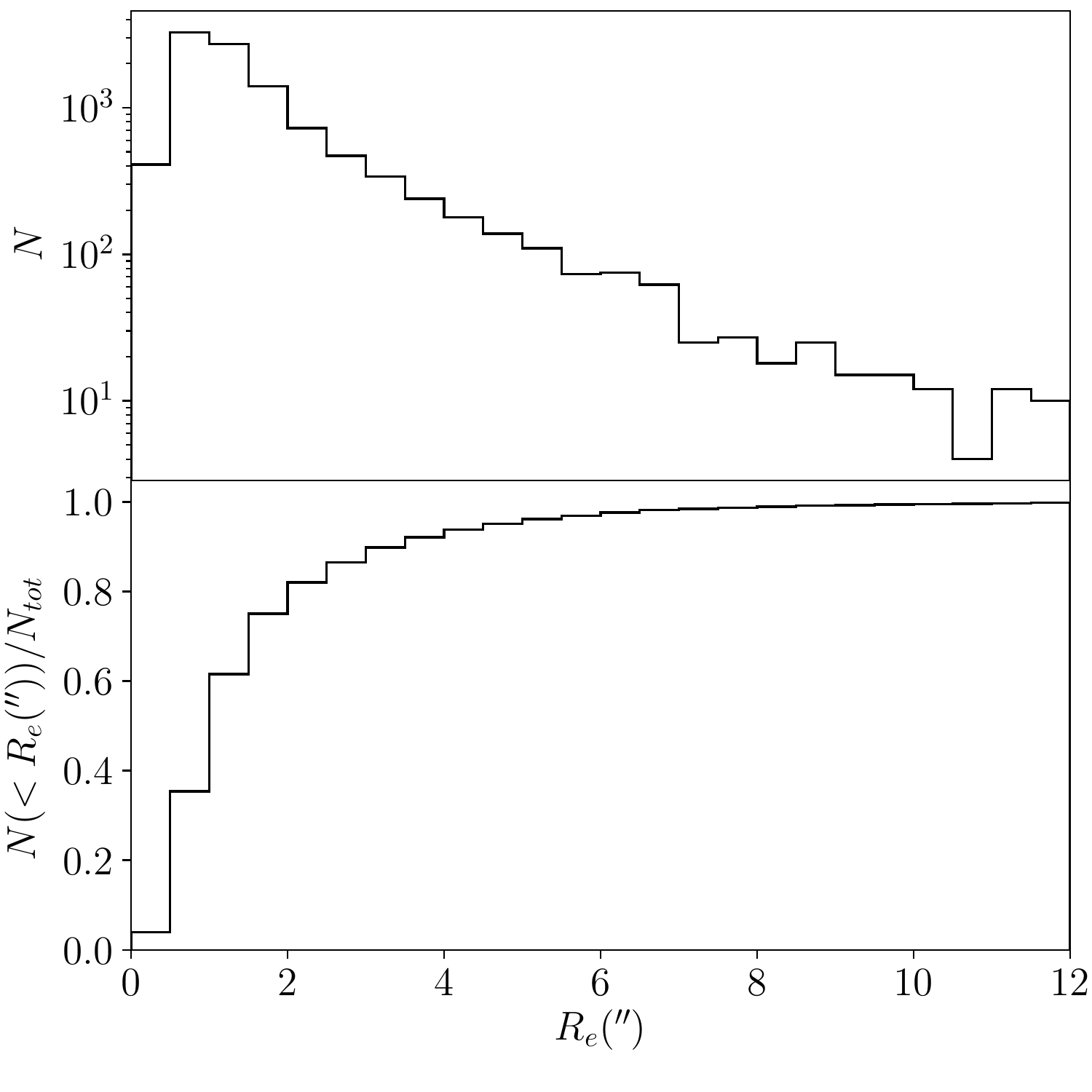}
 \caption{{\em Top:} Distribution in half-light radius, in angular units, of the galaxies in our sample. {\em Bottom:} cumulative distribution in half-light radius.}
 \label{fig:angreff}
\end{figure}

\Fref{fig:angreff} shows the distribution of half-light radii in angular units. For most objects, the half-light radius is well within the 8.4'' region used for the fit, with only 1\% of the galaxies in the sample exceeding this limit.

\begin{figure}
\centering
\includegraphics[width=\columnwidth]{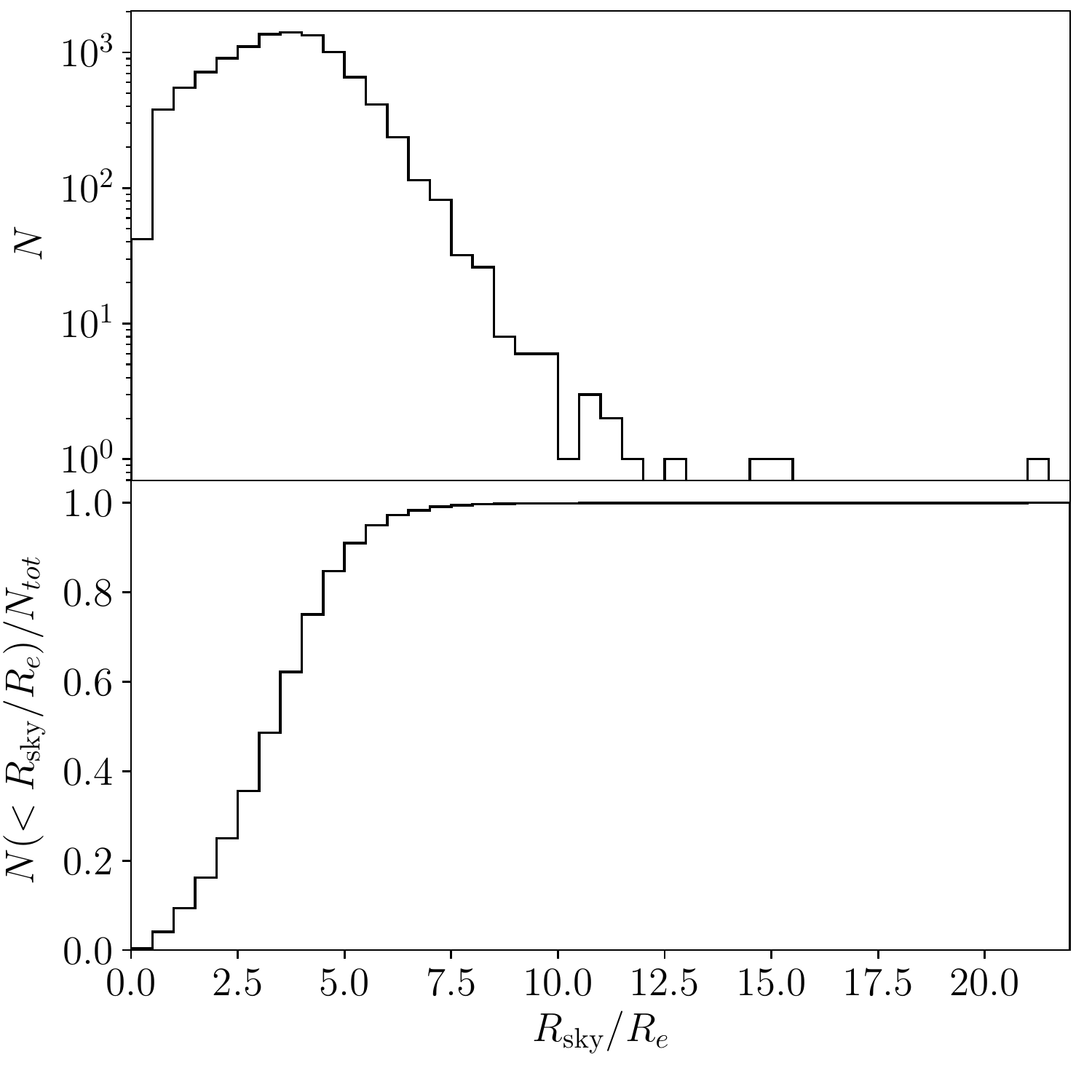}
 \caption{{\em Top:} distribution in the ratio between the radius at which the $i-$band surface brightness of a galaxy falls below the sky fluctuation level, $\rsky$, and the half light radius $\reff$. {\em Bottom:} cumulative distribution.}
 \label{fig:rsky}
\end{figure}

It is also interesting to check how far out from the centre of each galaxy the HSC data allows us to probe.
For this purpose, we compute, for each object, the radius at which the best-fit $i-$band S\'{e}rsic surface brightness falls below the level of the root mean square fluctuation from the sky background, $\rsky$.
We plot the distribution of $\rsky/\reff$ in \Fref{fig:rsky}. The median of this distribution is $3.6$, with only $4\%$ of the objects having a value of $\rsky/\reff$ smaller than unity.
For the typical galaxy in our sample, then, HSC $i-$band data allows us to detect flux out to $3.6$ times the half-light radius. For a de Vaucouleurs profile \citep[$n=4$][]{deV48}, the fraction of the total mass enclosed within this aperture is around $80\%$. This means that $\sim20\%$ (or $0.08$~dex) of the total flux of estimated for a typical galaxy is not directly observed, but rather the result of an extrapolation.

Around $5\%$ of the lenses are identified as brightest central galaxies (BCGs) in the HSC cluster catalogue of \citet{Ogu++18}. For these objects, we expect intra-cluster light to contribute with a non-negligible fraction to the observed surface brightness. We do not attempt to separate the light of the central galaxy from the intra-cluster light, as the distinction between these two components is not very well defined. For cluster BCGs, then, our measurements of the surface brightness profile, and the stellar mass and size derived from it, must be interpreted as that of the sum of the central galaxy and the intra-cluster light.

\subsection{Alternative models}\label{ssec:serexp}

Our fiducial model for the surface brightness distribution of our galaxies consists of a single S\'{e}rsic component, as described in subsection \ref{ssec:sersicprofile}.
In order to test for possible systematic effects related to this particular choice, we also consider two alternative models.
The first of these models consists of a single de Vaucouleurs profile: it is a particular case of the more general S\'{e}rsic profile, and is known to provide a good description of the average surface brightness profile of local early-type galaxies.
The second model is a two-component surface brightness profile, consisting of the sum of a S\'{e}rsic and an exponential profile (i.e. a S\'{e}rsic profile with $n=1$).
We impose that the two components have the same centroid and colours, but allow for different half-light radius, axis ratio and position angle between the two.
We will refer to this as the ``SerExp'' model throughout this paper.
The SerExp is a generalisation of the S\'{e}rsic model, and is expected to provide better fits in virtue of the higher number of degrees of freedom. 
\citet{Ber++14} showed that a SerExp model provides a better description of the surface brightness profile of SDSS galaxies, compared to models with a single S\'{e}rsic component.

\begin{figure*}
\includegraphics[width=\textwidth]{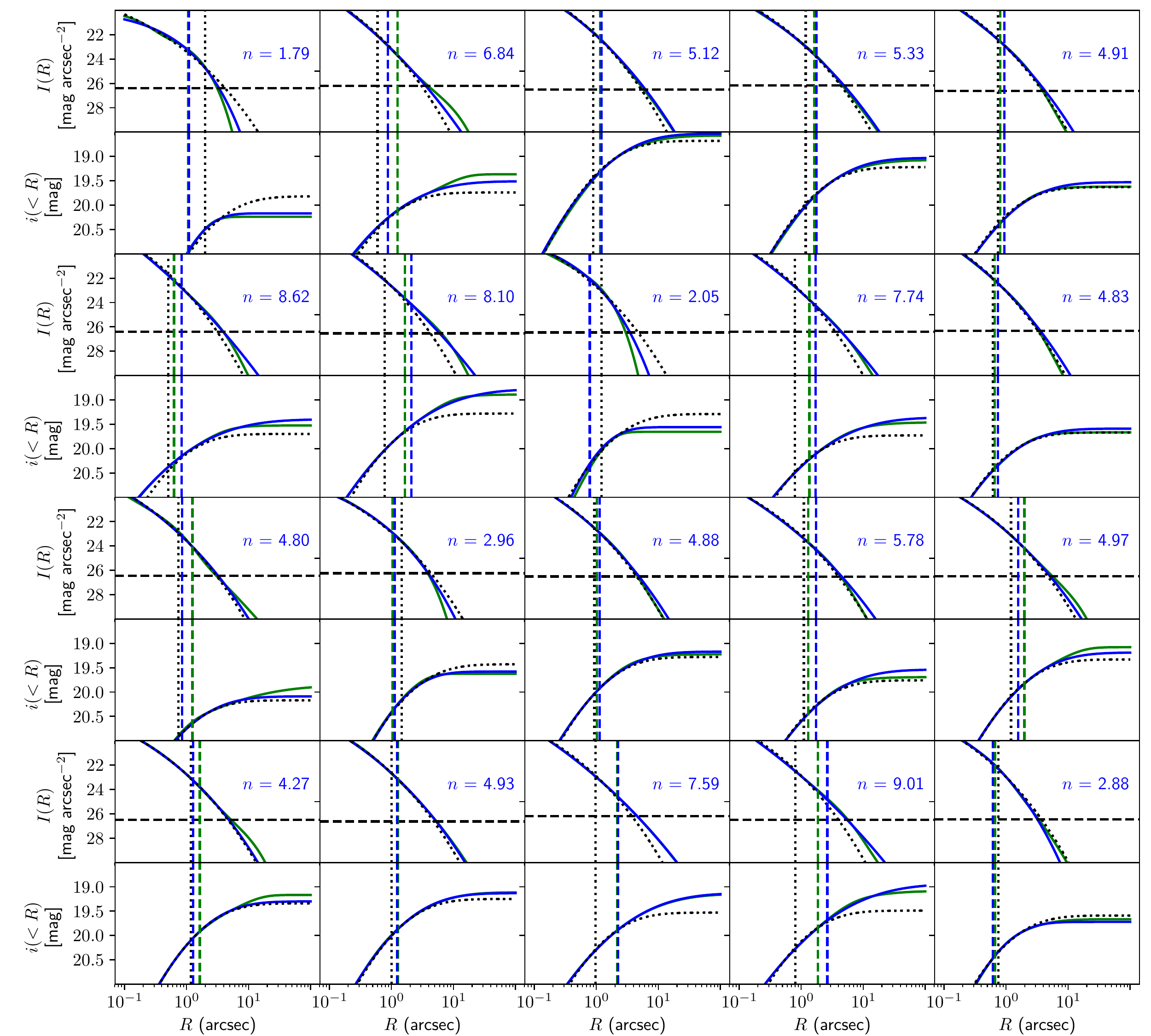}
\caption{$i-$band surface brightness profile (top sub-panel) and enclosed flux as a function of aperture (bottom sub-panel) for the best-fit S\'{e}rsic (blue), SerExp (green), and de Vaucouleurs (dotted black) profile for the 20 example objects shown in \Fref{fig:collage}.
The dashed horizontal line marks the surface brightness corresponding to the rms sky fluctuation. The dashed vertical lines mark the half-light radius.
}
\label{fig:profiles}
\end{figure*}

In \Fref{fig:profiles}, we show the circularised $i-$band surface brightness profiles as well as the enclosed flux as a function of aperture for 20 example galaxies (the same objects plotted in \Fref{fig:collage}), obtained for the S\'{e}rsic, the de Vaucouleurs and the SerExp model.
In the region probed by HSC data, roughly where the galaxy is brighter than the sky fluctuation level (marked by the horizontal dashed line in each subplot), the differences between the S\'{e}rsic (blue curves) and the SerExp model (green curves) are minimal. This is reassuring, as it means that the S\'{e}rsic model is sufficiently flexible to capture the complexity in the surface brightness profiles of our sample. 

A similar case can be made, for many objects, for the de Vaucouleurs model, although for some systems the data clearly prefers S\'{e}rsic indices different from $n=4$. These are the objects for which the best-fit de Vaucouleurs profile, shown as a dotted black curve, starts to deviate from the best-fit S\'{e}rsic profile well within the region above the background noise level. Typically, the residuals between the best-fit de Vaucouleurs profile and the data show under-subtraction at large radii, for the systems with $n > 4$, and over-subtraction for the systems with $n < 4$.

The difference between the large $R$ behaviour of the three models results in different total fluxes and, consequently, different half-light radii.
For most of the objects in our sample, our data does not allow us to decide which of the three models is more accurate, especially between the S\'{e}rsic and the SerExp: the differences arise at large radii, where the surface brightness is too faint to be detected by HSC data.
We will then carry out the analysis with all models, and check whether the results are stable with respect to the particular assumption on the large $R$ behaviour of the surface brightness profile of our galaxies.

Although the de Vaucouleurs model produces bad fits for a non-negligible fraction of the CMASS galaxies in our sample, we still show results obtained with this particular surface brightness profile choice, since many studies in the literature are based on the same model.

\section{Generalised mass-size relation}\label{sect:masssize}

\subsection{Basic model}\label{ssec:mr_rel}

We fit for the mass-size relation of our sample of massive ($\log{\mobs} > 11$) elliptical galaxies using a Bayesian hierarchical approach:
we model the distribution of true stellar mass and size of the sample with a functional form, described by a set of {\em hyper-parameters} $\hyperp$, which we then infer from the data.
We choose the following form for this distribution:
\begin{equation}\label{eq:hierarch}
\pr(\mstar,\reff|\hyperp) = \mathcal{S}(M_*)\mathcal{R}(\reff|\mstar).
\end{equation}
Here $\mathcal{S}(M_*)$ is a skew Gaussian in $\log{\mstar}$:
\begin{equation}\label{eq:fullskew}
\mathcal{S}(\mstar) = \frac{1}{\sqrt{2\pi}\sigma_*}\exp{\left\{-\frac{(\log{\mstar} - \mu_*)^2}{2\sigma_*^2}\right\}}\Phi(\log{\mstar}),
\end{equation}
with
\begin{equation}\label{eq:skew}
\Phi(\log{\mstar}) = 1 + \mathrm{erf}\left(s_*\frac{\log{\mstar} - \mu_*}{\sqrt{2}\sigma_*}\right),
\end{equation}
and $\mathcal{R}(\reff|\mstar)$ is a Gaussian in $\log{\reff}$ 
\begin{equation}\label{eq:redist}
\mathcal{R}(\reff|\mstar) = \frac{1}{\sqrt{2\pi}\sigma_R}\exp{\left\{-\frac{(\log{\reff} - \mu_R(\mstar))^2}{2\sigma_R^2}\right\}},
\end{equation}
with a mean that scales with stellar mass as follows
\begin{equation}\label{eq:mure}
\mu_R(\mstar) = \mu_{R,0} + \beta_R(\log{\mstar} - 11.4).
\end{equation}
The hyper-parameters describing this 2-dimensional distribution are:
\begin{equation}
\hyperp \equiv \{\mu_*, \sigma_*, s_*, \mu_{R,0}, \sigma_R, \beta_R\}.
\end{equation}
$\mu_*$, $\sigma_*$ and $s_*$ describe the stellar mass distribution. If the skewness parameter $s_*$ is set to 0, this reduces to a Gaussian with mean $\mu_*$ and $\sigma_*$. Positive values of $s_*$ correspond to a distribution with a long tail above $\mu_*$ and a sharper cutoff at low masses.
The parameter $\mu_{R,0}$ is the average value of $\log{\reff}$ at the pivot stellar mass $\log{\mstar}=11.4$, $\sigma_*$ is the scatter in $\log{\reff}$ at fixed stellar mass, and finally $\beta_R$ is a power-law dependence of $\reff$ on $\mstar$.

Our goal is to infer the posterior probability distribution of the set of hyper-parameters $\hyperp$ given the data $\data$. Using Bayes' theorem,
\begin{equation}\label{eq:bayes}
\pr(\hyperp|\data) \propto \pr(\hyperp)\pr(\data|\hyperp),
\end{equation}
where $\pr(\hyperp)$ is the prior on the hyper-parameters and $\pr(\data|\hyperp)$ is the likelihood of observing the data given the value of the hyper-parameters.
Assuming measurements on separate galaxies are independent from each other, the latter is the following product over individual objects:
\begin{equation}\label{eq:likeint}
\pr(\data|\hyperp) = \prod_i \int d\mstari d\reffi \pr(\datai|\mstari,\reffi)\pr(\mstari,\reffi|\hyperp).
\end{equation}
In the above equation we have introduced a set of latent variables, the true values of the stellar mass and half-light radius of each galaxy. 
These are necessary to calculate the likelihood. 
We are not interested in constraining the stellar mass and size of individual galaxies, so we marginalise over all possible values, modulated by $\pr(\mstari,\reffi|\hyperp)$, which is the distribution introduced in \Eref{eq:hierarch}.

The data consists of the observed values of the stellar mass and the half-light radius, $\mobs$ and $\robs$, and related uncertainties.
For simplicity, we neglect the uncertainty on size, since it is on the order of $2\%$. 
As a result, the covariance between the measurement of the half-light radius and the stellar mass is also set to zero. This is a fair approximation, since the uncertainty on the stellar mass is dominated by systematic uncertainties in the stellar population synthesis model.
The first term in the integrand of \Eref{eq:likeint} then becomes
\begin{equation}
\pr(\datai|\mstari,\reffi) = \pr(\mobsi|\mstari)\delta(\robsi - \reffi),
\end{equation}
and \Eref{eq:likeint} reduces to
\begin{equation}\label{eq:likeint2}
\pr(\data|\hyperp) = \prod_i \int d\mstari \pr(\mobsi|\mstari)\pr(\mstari,\reffi|\hyperp).
\end{equation}
We approximate the likelihood of obtaining a value of the observed stellar mass $\mobsi$ given the true value $\mstari$ as a Gaussian in $\log{\mstari}$:
\begin{equation}\label{eq:likemobs}
\pr(\mobsi|\mstari) = \frac{A_i}{\sqrt{2\pi}\epsilon_{*,i}}\exp{\left\{-\frac{(\log{\mstari} - \log{\mobsi})^2}{2\epsilon_{*,i}}\right\}}.
\end{equation}
In the equation above, $\epsilon_{*,i}$ is the standard deviation in $\log{\mstari}$, as obtained from the MCMC chain of the stellar population fit, and $A_i$ is a normalisation constant that ensures that the integral of the likelihood over all possible values of $\mobsi$ is one. Since we applied the cut $\log{\mobs} > 11.0$ to define our sample, this is equivalent to
\begin{equation}
\int_{11.0}^\infty d\log{\mobsi} \frac{A_i}{\sqrt{2\pi}\epsilon_{*,i}}\exp{\left\{-\frac{(\log{\mstari} - \log{\mobsi})^2}{2\epsilon_{*,i}}\right\}} = 1.
\end{equation}

\begin{figure}
\centering
\includegraphics[width=\columnwidth]{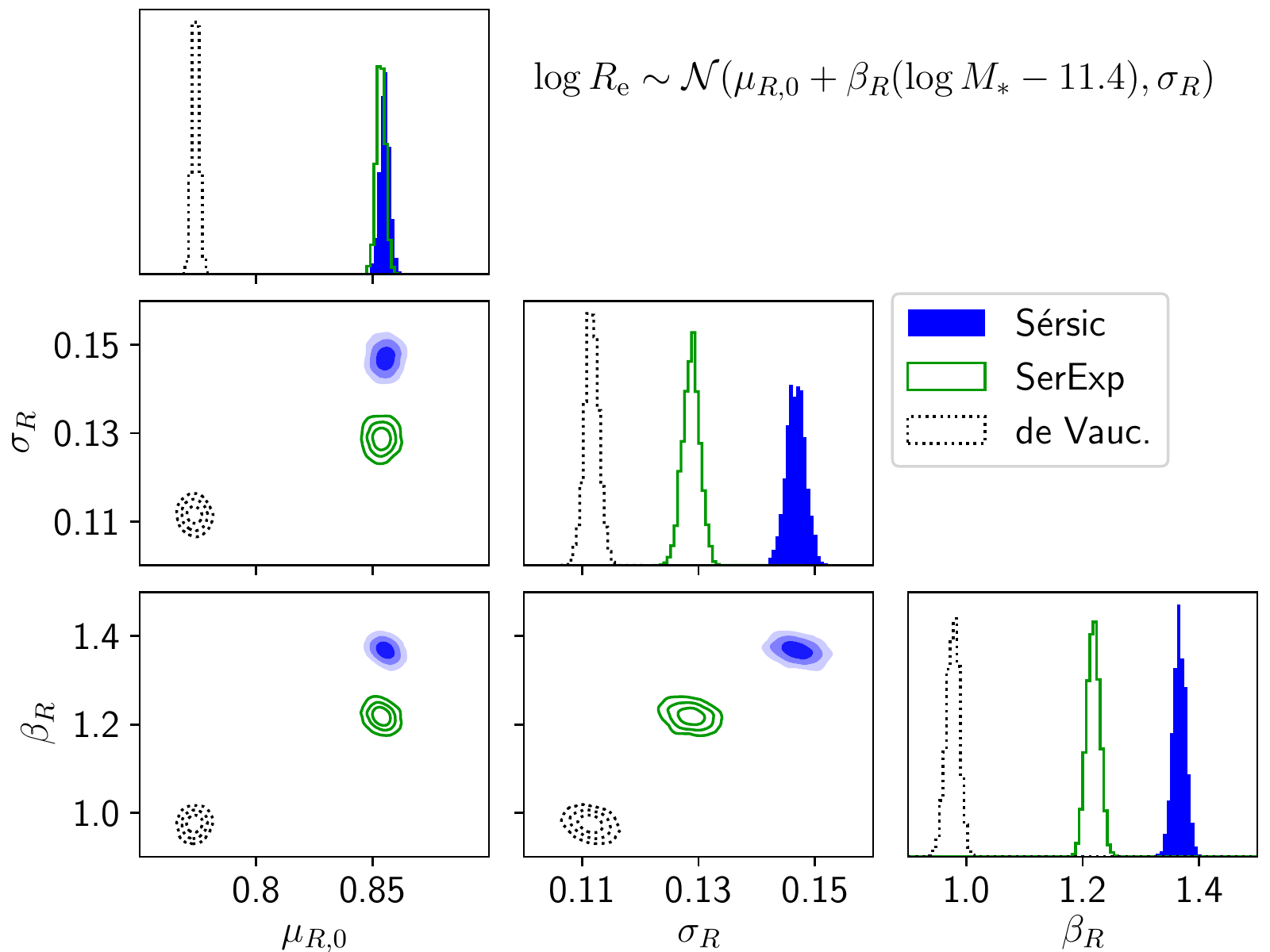}
 \caption{Posterior probability distribution of the parameters describing the distribution of galaxy sizes as a function of stellar mass, introduced in subsection \ref{ssec:mr_rel}. Results based on the S\'{e}rsic, SerExp and de Vaucouleurs model are shown.}
 \label{fig:mrcp}
\end{figure}

We sample the posterior probability distribution of the hyper-parameters $\hyperp$ given the data $\data$, \Eref{eq:bayes}, by running an MCMC.
We calculate the integrals in \Eref{eq:likeint2} with the importance sampling and Monte Carlo integration method described in SL18.
We assume flat priors on all hyper-parameters except $s_*$, for which we assume a flat prior on its $10$-base logarithm.
In \Fref{fig:mrcp} we plot the posterior probability distribution relative to the parameters describing the distribution of sizes: $\mu_{R,0}$, $\sigma_R$ and $\beta_R$.
The median and 68\% credible interval of all hyper-parameters is reported in \Tref{tab:mr}.
We show results based on the S\'{e}rsic, SerExp and de Vaucouleurs models.

We find a remarkably steep mass-size relation, although the exact value of the dependence of $\reff$ on $\mstar$ depends on the choice of the photometric model: we obtain $\beta_R=1.37\pm0.01$ if the S\'{e}rsic photometric model is used and $\beta_R=1.22\pm0.01$ in the SerExp case.
The difference between the models appears to be driven by differences in size at the high mass end of the distribution: at fixed stellar mass, the values of $\reff$ obtained with the SerExp model are slightly smaller compared to the S\'{e}rsic values, as can be seen from the bottom-left panel of \Fref{fig:mass_size}. 
A similar behaviour was found by \citet{Ber++14} on SDSS galaxies.
As a result, the uncertainty on our inference on the slope of the mass-size relation is dominated by systematics related to the particular choice of the surface brightness profile used to describe CMASS galaxies.

The difference between the S\'{e}rsic and the de Vaucouleurs sizes is much more evident, hence the much lower value of the mass-size relation slope inferred for this particular model: $\beta_R=0.98\pm0.01$. However, we believe this to be a biased inference, since a de Vaucouleurs model provides a poor description of the surface brightness profile of a non-negligible fraction of the objects in our sample.

\begin{table*}
\caption{Mass-size relation, as modelled in subsection \ref{ssec:mr_rel}. Median values and 68\% credible interval of the posterior probability distribution of individual hyper-parameters, marginalised over the rest of the hyper-parameters. Results based on the S\'{e}rsic, SerExp and de Vaucouleurs model are reported.
}
\label{tab:mr}
\begin{tabular}{lcccl}
\hline
\hline
 & S\'{e}rsic model & SerExp model & de Vauc. model & Parameter description \\
\hline
$\mu_{R,0}$ & $0.855 \pm 0.002$ & $0.854 \pm 0.002$ & $0.774 \pm 0.002$ & Average $\log{\reff}$ at stellar mass $\log{\mstar}=11.4$\\
$\sigma_R$ & $0.147 \pm 0.002$ & $0.129 \pm 0.001$ & $0.112 \pm 0.001$ & Dispersion in $\log{\reff}$ around the average\\
$\beta_R$ & $1.366 \pm 0.011$ & $1.218 \pm 0.011$ & $0.977 \pm 0.011$ & Power-law dependence of size on $\mstar$\\
$\mu_*$ & $11.249 \pm 0.004$ & $11.274 \pm 0.005$ & $11.252 \pm 0.006$ & Mean-like parameter of Gaussian term in \Eref{eq:fullskew}\\
$\sigma_*$ & $0.285 \pm 0.004$ & $0.254 \pm 0.004$ & $0.202 \pm 0.004$ & Dispersion-like parameter of Gaussian term in \Eref{eq:fullskew}\\
$\log{s_*}$ & $0.44 \pm 0.02$ & $0.31 \pm 0.02$ & $0.17 \pm 0.03$ & Log of the skewness parameter in \Eref{eq:skew}\\

\hline
\end{tabular}
\end{table*}

\subsection{Distribution of size and S\'{e}rsic index}\label{ssec:mnr_rel}

We add complexity to our model by considering the distribution in S\'{e}rsic index of our galaxies, in addition to stellar mass and size. 
We model the distribution in $\mstar$, $\nser$ and $\reff$ as follows:
\begin{equation}\label{eq:mnrdist}
\pr(\mstar, \nser, \reff | \hyperp) = \mathcal{S}(M_*)\mathcal{I}(\nser|M_*)\mathcal{R}(\reff|\mstar,\nser).
\end{equation}
The term $\mathcal{S}(M_*)$ is the same introduced in \Eref{eq:fullskew} and \Eref{eq:skew}. $\mathcal{I}(\nser|M_*)$ is a Gaussian distribution in the base-10 logarithm of the S\'{e}rsic index,
\begin{equation}\label{eq:nserdist}
\mathcal{I}(\nser|M_*) = \frac{1}{\sqrt{2\pi}\sigma_\nser}\exp{\left\{-\frac{(\log{\nser} - \mu_\nser(\mstar))^2}{2\sigma_\nser^2}\right\}},
\end{equation}
with a mean that scales with stellar mass as
\begin{equation}\label{eq:nsermu}
\mu_\nser(\mstar) = \mu_{\nser,0} + \beta_\nser(\log{\mstar} - 11.4)
\end{equation}
and dispersion $\sigma_\nser$.
Finally, we keep the same form as \Eref{eq:redist} for the term describing the distribution in half-light radius, $\mathcal{R}$, but update \Eref{eq:mure} by adding a dependence on the S\'{e}rsic index to the average size:
\begin{equation}\label{eq:reffnsermu}
\mu_R(\mstar) = \mu_{R,0} + \beta_R(\log{\mstar} - 11.4) + \nu_R\log{\nser/4}.
\end{equation}
At fixed stellar mass, \Eref{eq:mnrdist} is a bi-variate Gaussian in $\log{\nser}$ and $\log{\reff}$.

\begin{figure*}
\centering
\includegraphics[width=\textwidth]{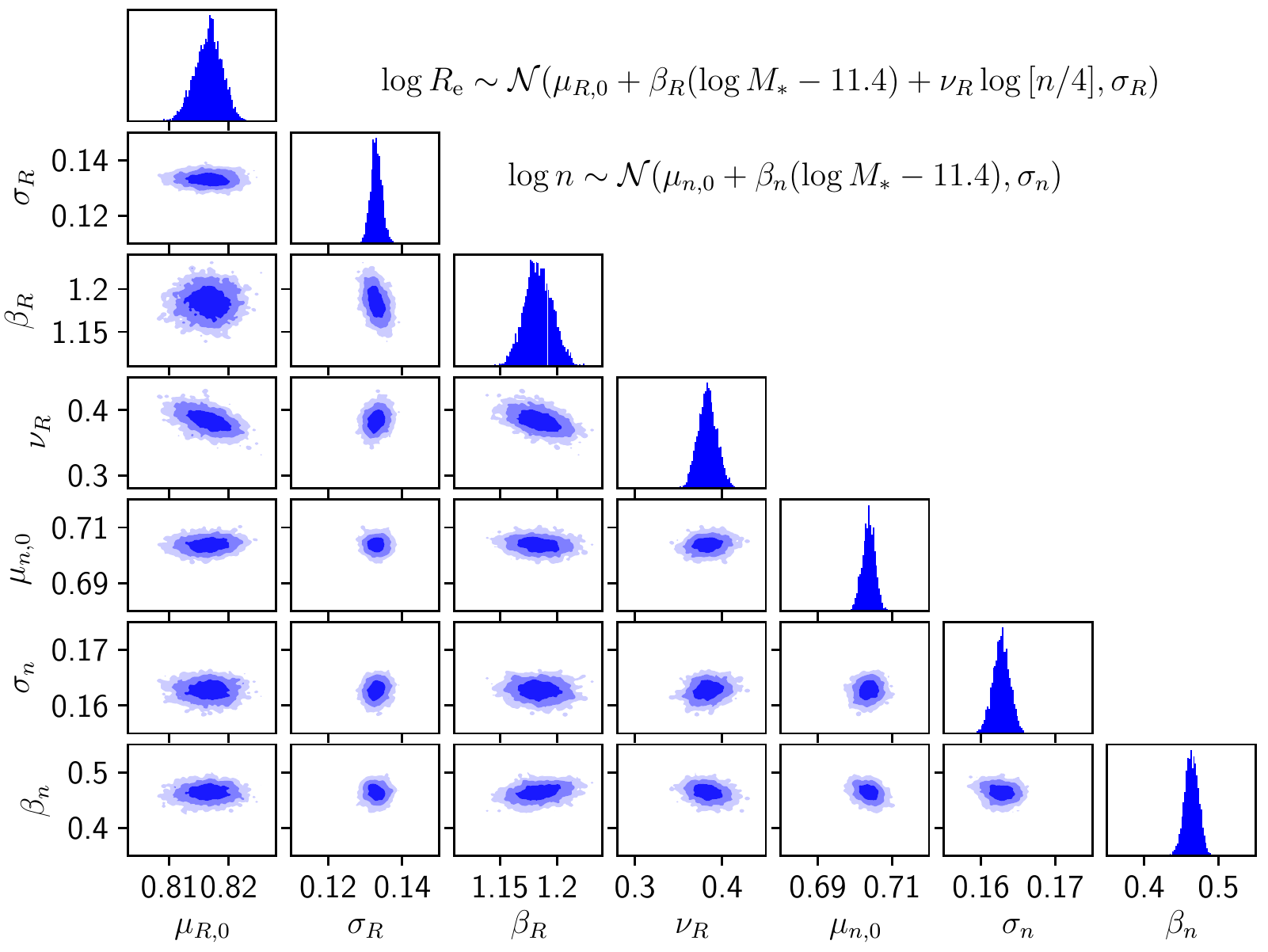}
 \caption{Posterior probability distribution of the parameters describing the distribution of S\'{e}rsic index and galaxy size.}
 \label{fig:mnrcp}
\end{figure*}

As we have done for the sizes, we assume that the S\'{e}rsic indices are measured exactly, since the uncertainties on $\nser$ are very small.
In \Fref{fig:mnrcp} we plot the posterior probability distribution of the hyper-parameters describing the distribution in S\'{e}rsic index, $\mu_{\nser,0}$, $\sigma_{\nser}$ and $\beta_\nser$, as well as the hyper-parameters describing the distribution in size, including the new hyper-parameter $\nu_R$, which describes the dependence of size on $\nser$.
The median and 68\% credible values of all hyper-parameters are listed in \Tref{tab:mnr}.
We show only result based on the S\'{e}rsic model, as $\nser$ is not well defined in the case of a SerExp profile and is fixed in the case of a de Vaucouleurs model.
\begin{table*}
\caption{Mass-S\'{e}rsic index-size relation, as modelled in subsection \ref{ssec:mnr_rel}. Median values and 68\% credible interval of the posterior probability distribution of individual hyper-parameters, marginalised over the rest of the hyper-parameters.
}
\label{tab:mnr}
\begin{tabular}{lcl}
\hline
\hline
 & & Parameter description \\
\hline
$\mu_{\nser,0}$ & $0.704 \pm 0.002$ & Average $\log{\nser}$ at stellar mass $\log{\mstar}=11.4$\\
$\sigma_{\nser}$ & $0.163 \pm 0.001$ & Dispersion in $\log{\nser}$ around the average\\
$\beta_{\nser}$ & $0.464 \pm 0.009$ & Power-law dependence of S\'{e}rsic index on $\mstar$\\
$\mu_{R,0}$ & $0.817 \pm 0.002$ & Average $\log{\reff}$ at stellar mass $\log{\mstar}=11.4$ and S\'{e}rsic index $\nser=4$\\
$\sigma_R$ & $0.133 \pm 0.002$ & Dispersion in $\log{\reff}$ around the average\\
$\beta_R$ & $1.184 \pm 0.012$ & Power-law dependence of size on $\mstar$\\
$\nu_R$ & $0.383 \pm 0.011$ & Power-law dependence of size on $\nser$\\
$\mu_*$ & $11.249 \pm 0.004$ & Mean-like parameter of Gaussian term in \Eref{eq:fullskew}\\
$\sigma_*$ & $0.285 \pm 0.004$ & Dispersion-like parameter of Gaussian term in \Eref{eq:fullskew}\\
$\log{s_*}$ & $0.43 \pm 0.02$ & Log of the skewness parameter in \Eref{eq:skew}\\

\hline
\end{tabular}
\end{table*}

The inferred average value of $\log{\nser}$ at the pivot stellar mass $\log{\mstar}=11.4$ is $\mu_{\nser,0} = 0.704\pm0.002$, corresponding to a S\'{e}rsic index $\nser \approx 5.1$.
We find a positive correlation between S\'{e}rsic index and stellar mass, quantified by the parameter $\beta_\nser = 0.464\pm0.009$.
This is closely related to the well-known S\'{e}rsic index - luminosity correlation \citep{CCD93}, and tells us that, for instance, the average S\'{e}rsic index of galaxies with stellar mass $\log{\mstar}=11.8$ is as high as $\nser\approx7.8$. 
We also find a correlation between size and S\'{e}rsic index {\em at fixed stellar mass}: $\nu_R = 0.38\pm0.01$.

The value of parameter $\beta_R$, the correlation between mass and size, is smaller compared to the value obtained in the analysis of subsection \ref{ssec:mr_rel}, based on the simpler model ($1.18$ vs $1.37$).
This is because, in the context of this more complex model, $\beta_R$ quantifies how size scales with stellar mass {\em at fixed S\'{e}rsic index}. 
The combination of this scaling with 1) the dependence of size on $\nser$ (positive value of parameter $\nu_R$) and 2) the positive correlation between S\'{e}rsic index and stellar mass, produces the steep mass-size relation observed.

At the same time, the inferred value for the average size at the pivot point, $\mu_{R,0}$, is smaller than the value obtained previously. This is because $\mu_{R,0}$ now refers to the average $\log{\reff}$ at a stellar mass $\log{\mstar}=11.4$, and S\'{e}rsic index $\nser=4$, corresponding to a de Vaucouleurs profile, which is a smaller value of $\nser$ with respect to the average for galaxies of that mass.

The high S\'{e}rsic indices measured in this study provide a good description of the surface brightness profile of CMASS galaxies in the region probed by the data, which roughly corresponds to the radial range $2\rm{kpc} \lesssim R \lesssim 30\rm{kpc}$. The lower limit is set by the finite resolution of HSC data. An extrapolation of the best-fit S\'{e}rsic profile to the very inner regions produces very steep inner surface brightness profiles. Such cuspy profiles are typically not observed in the centres of nearby massive galaxies, but cannot be ruled out in our observations, due to the atmospheric blurring of the images. The impact of a potentially inaccurate description of the inner surface brightness profile on the stellar mass and size measurement used for our study is nevertheless minimal: the mass enclosed in the region significantly affected by the atmospheric seeing is only a small fraction of the total.
However, we urge caution when using our derived S\'{e}rsic fits to predict quantities that are more sensitive to the inner stellar distribution, such as the dynamical mass.

The outer limit to the region where the S\'{e}rsic profile is most accurate is set by the radius at which the surface brightness of CMASS galaxies falls below the noise level of HSC data. The median value of this radius for our sample is $28\rm{kpc}$. 

\section{Generalised stellar-to-halo mass relation}\label{sect:halos}

We now use weak lensing measurements from HSC to obtain the dark matter halo mass distribution of our sample, as a function of galaxy properties.
We explore two different models, with increasing degrees of complexity.
In the first model, presented in the next subsection, we let halo mass to scale with stellar mass and half-light radius. In subsection \ref{ssec:halonser}, we generalise this model to allow for a possible additional dependence of halo mass on S\'{e}rsic index.

Mirroring the approach adopted in \Sref{sect:masssize}, we will fit the simpler of the two models to data obtained with the three different parameterisations of the surface brightness profiles, in order to explore the impact of potential systematics associated with the photometry fitting on the derived stellar-to-halo mass relation.
We will then fit the full model to our fiducial S\'{e}rsic profile-based measurements.

\subsection{Halo mass as a function of stellar mass and size}\label{ssec:halosize}

We will start by considering the joint distribution in stellar mass $\mstar$, half-light radius $\reff$ and halo mass $\mhalo$.
Similarly to the approach adopted in the previous Section, we model this distribution as follows:
\begin{equation}\label{eq:fulldist}
\pr(\mstar,\reff,\mhalo) = \mathcal{S}(\mstar)\mathcal{R}(\reff|\mstar)\mathcal{H}(\mhalo|\mstar,\reff).
\end{equation}
The terms $\mathcal{S}$ and $\mathcal{R}$ are the same one introduced in subsection \ref{ssec:mr_rel}: a skew Gaussian in $\log{\mstar}$ and a Gaussian in $\log{\reff}$ with a mean that scales with stellar mass.
The term relative to the halo mass, $\mathcal{H}$, is modelled as a Gaussian in $\log{\mhalo}$,
\begin{equation}\label{eq:haloterm}
\mathcal{H}(\mhalo|M_*,\reff) = \frac{1}{\sqrt{2\pi}\sigma_h}\exp{\left\{-\frac{(\log{\mhalo} - \mu_h(\mstar,\reff))^2}{2\sigma_h^2}\right\}},
\end{equation}
with a mean that scales with stellar mass and size as
\begin{equation}\label{eq:muhalo}
\mu_h(\mstar,\reff) = \mu_{h,0} + \beta_h(\log{\mstar} - 11.4) + \xi_h(\log{\reff} - \mu_R(\mstar)),
\end{equation}
and dispersion $\sigma_h$.
The term $\mu_R(\mstar)$ in \Eref{eq:muhalo} is the average $\log{\reff}$ for galaxies of mass $\mstar$, introduced in \Eref{eq:mure}. The parameter $\xi_h$, then, is a power-law dependence of halo mass on ''excess size'', defined as the ratio between the half-light radius of a galaxy and the average size for galaxies of the same stellar mass.
This excess size is conceptually similar to the ``mass-normalised radius'' used by \citet{New++12} and \citet{Hue++13}.

The full list of hyper-parameters of the model is then
\begin{equation}
\hyperp \equiv \{\mu_{h,0}, \sigma_h, \beta_h, \xi_h, \mu_*, \sigma_*, s_*, \mu_{R,0}, \sigma_R, \beta_R\}.
\end{equation}
In order to infer the posterior probability distribution of the hyper-parameters given the data, we need to be able to evaluate the likelihood $\pr(\data|\hyperp)$ for any value of $\hyperp$.
Under the isolated lens assumption, on which the S18 weak lensing method is based, the likelihood of the data relative to each lens is independent from each other. We can then write, analogously to \Eref{eq:likeint},
\begin{equation}
\pr(\data|\hyperp) = \prod_i \pr(\datai|\hyperp).
\end{equation}
For each galaxy, the data consists of the measured stellar mass, effective radius, and shape measurements of background sources within a cone of $300$ physical kpc radius at the redshift of the lens (as previously discussed in subsection \ref{ssec:wldata}).
In order to evaluate the likelihood of the shape measurements, it is necessary to assume a model for the mass distribution of the lens.
We first define the halo mass as the dark matter mass enclosed in a shell with average density equal to 200 times the critical density of the Universe, commonly referred to as $M_{200}$.
We then {\em assume a spherical Navarro Frenk and White \citep[NFW][]{NFW97} shape} for the dark matter halo density profile:
\begin{equation}
\rho(r) \propto \frac{1}{r(1+r/r_s)^2}.
\end{equation}
Finally, we assume a mass-concentration relation from \citet{Mac++08}: given $r_{200}$ (the radius enclosing a mass equal to $M_{200}$), we assume that the concentration $c_h = r_{200}/r_s$ is drawn from the following Gaussian in its base-10 logarithm:
\begin{equation}\label{eq:mcrel}
\pr(c_h|\mhalo) = \frac{1}{\sqrt{2\pi}\sigma_c}\exp{\left\{-\frac{(\log{c_h} - \mu_c(\mhalo))^2}{2\sigma_c}\right\}},
\end{equation}
with mean $\mu_c$ that scales with halo mass as
\begin{equation}
\mu_c(\mhalo) = \mu_{c,0} + \beta_c(\log{\mhalo} - 12)
\end{equation}
and dispersion $\sigma_c$.
We fix $\mu_{c,0} = 0.830$, $\beta_c=-0.098$ and $\sigma_c=0.10$.

For each galaxy, the likelihood of observing the data given the value of the hyper-parameters is obtained by marginalising over all possible values of the concentration, as well as stellar and halo mass and half-light radius:
\begin{equation}
\begin{split}
\pr(\datai|\hyperp) = & \int d\mstari d\mhaloi dc_{h,i} d\reffi \pr(\datai|\mstari,\mhaloi, c_{h,i},\reffi) \times \\
& \pr(c_{h,i}|\mhaloi)\pr(\mstari,\mhaloi,\reffi|\hyperp).
\end{split}
\end{equation}
Here $\pr(c_{h,i}|\mhaloi)$ is the mass-concentration relation introduced in \Eref{eq:mcrel}, which acts as a prior on the concentration parameter.
As done throughout \Sref{sect:masssize}, we set the uncertainty on the half-light radius to zero, so that the integral over $\reff$ reduces to that over a delta function.
The likelihood of the weak lensing data is obtained by evaluating the shear produced by the lens, given the values of $\mstar$, $\mhalo$ and $c_h$, corrected by an additive and multiplicative bias on the measurements, obtained via simulations by \citet{Man++18}. We refer to SL18 and \citet[][subsection 3.4]{Son++18} for more details on the calculation of the likelihood term for weak lensing data in the SL18 formalism.

\begin{figure*}
\centering
\includegraphics[width=\textwidth]{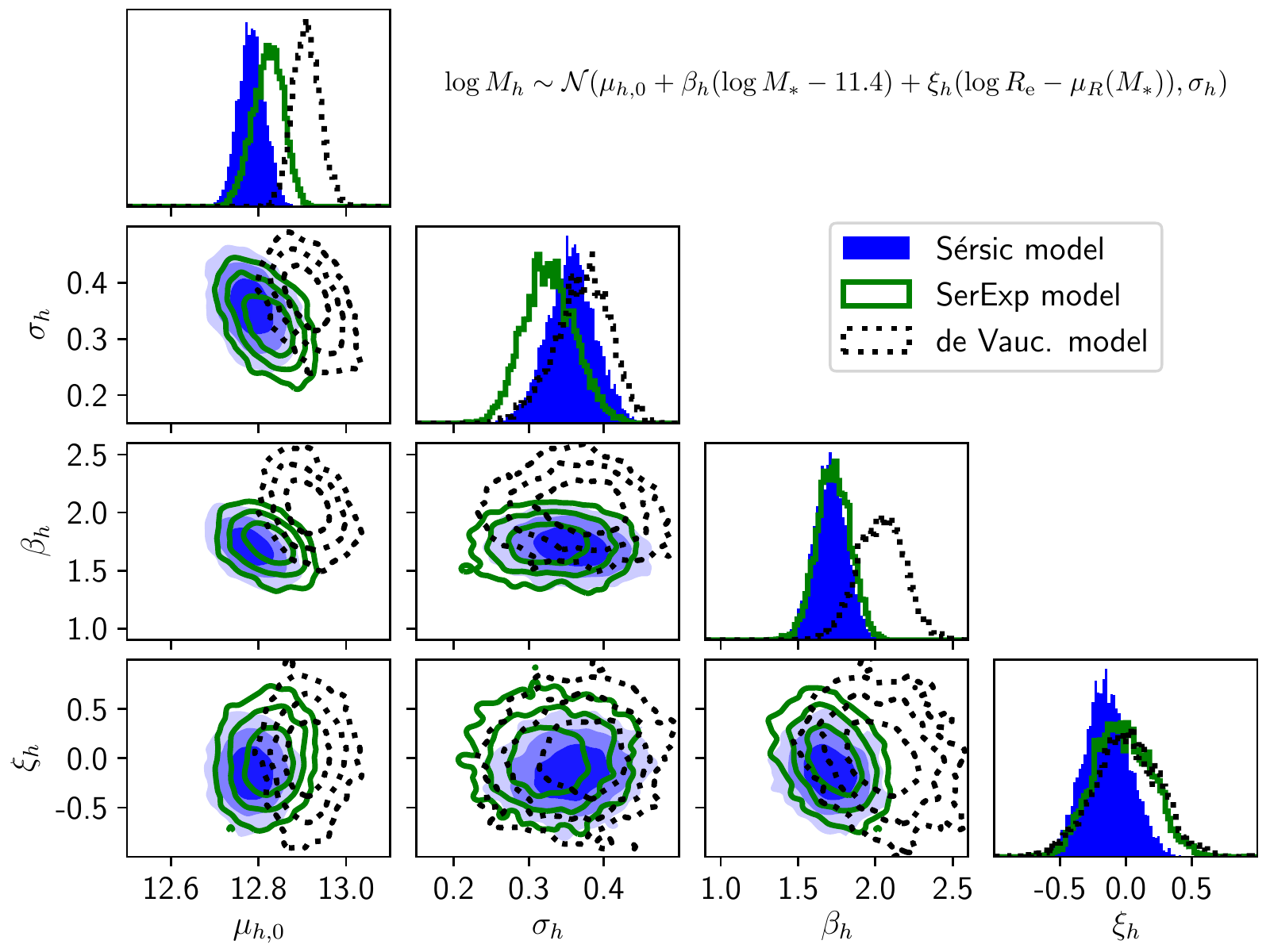}
 \caption{Posterior probability distribution of the parameters describing the distribution of halo mass, for the model introduced in subsection \ref{ssec:halosize}. In particular, $\mu_{h,0}$ is the average $\log{\mhalo}$ at a stellar mass $\log{\mstar}=11.4$ and average size for that mass, $\sigma_h$ is the dispersion in $\log{\mhalo}$ around the average, $\beta_h$ is a power-law scaling of halo mass with stellar mass, and $\xi_h$ is a power-law scaling of halo mass with excess size (defined as the ratio between the size of a galaxy and the average size of galaxies of the same stellar mass).}
 \label{fig:halocp}
\end{figure*}

In \Fref{fig:halocp}, we plot the posterior probability distribution of the hyper-parameters belonging to the halo mass term $\mathcal{H}$ in \Eref{eq:fulldist}. The median and 68\% credible interval on the full set of hyper-parameters is given in \Tref{tab:halo}.
We show results based on the S\'{e}rsic, SerExp and de Vaucouleurs model.
\begin{table*}
\caption{Median values and 68\% credible interval of the posterior probability distribution of the hyper-parameters describing the distribution in stellar mass, half-light radius and halo mass of the galaxies in our sample, as inferred by fitting the model introduced in subsection \ref{ssec:halosize}.
}
\label{tab:halo}
\begin{tabular}{lcccl}
\hline
\hline
 & S\'{e}rsic model & SerExp model & de Vauc. model & Parameter description \\
\hline
$\mu_{h,0}$ & $12.79 \pm 0.03$ & $12.83 \pm 0.03$ & $12.91 \pm 0.03$ & Average $\log{M_h}$ at stellar mass $\log{\mstar}=11.4$ and average size\\
$\sigma_h$ & $0.35 \pm 0.03$ & $0.32 \pm 0.03$ & $0.37 \pm 0.04$ & Dispersion in $\log{M_h}$ around the average\\
$\beta_h$ & $1.70 \pm 0.10$ & $1.73 \pm 0.11$ & $2.04 \pm 0.15$ & Power-law dependence of halo mass on $\mstar$ at fixed size\\
$\xi_h$ & $-0.14 \pm 0.17$ & $-0.03 \pm 0.21$ & $0.00 \pm 0.25$ & Power-law dependence of halo mass on excess size at fixed $M_*$\\
$\mu_{R,0}$ & $0.854 \pm 0.002$ & $0.852 \pm 0.002$ & $0.775 \pm 0.002$ & Average $\log{\reff}$ at stellar mass $\log{\mstar}=11.4$\\
$\sigma_R$ & $0.140 \pm 0.002$ & $0.124 \pm 0.002$ & $0.108 \pm 0.001$ & Dispersion in $\log{\reff}$ around the average\\
$\beta_R$ & $1.390 \pm 0.012$ & $1.239 \pm 0.011$ & $1.004 \pm 0.011$ & Power-law dependence of size on $\mstar$\\
$\mu_*$ & $11.246 \pm 0.004$ & $11.271 \pm 0.004$ & $11.249 \pm 0.005$ & Mean-like parameter of Gaussian term in \Eref{eq:fullskew}\\
$\sigma_*$ & $0.286 \pm 0.004$ & $0.256 \pm 0.004$ & $0.203 \pm 0.004$ & Dispersion-like parameter of Gaussian term in \Eref{eq:fullskew}\\
$\log{s_*}$ & $0.47 \pm 0.02$ & $0.34 \pm 0.02$ & $0.20 \pm 0.03$ & Log of the skewness parameter in \Eref{eq:skew}\\

\hline
\end{tabular}
\end{table*}

The average $\log{\mhalo}$ for galaxies of $\log{\mstar}=11.4$ and average size for their mass is $\mu_{h,0}=12.79\pm0.03$. We then infer a positive correlation between halo mass and stellar mass at fixed size, $\beta_h=1.70\pm0.10$. However, we do not find any strong evidence for an additional correlation between halo mass and size at fixed stellar mass. The median and $1\sigma$ limit of the parameter describing the dependence on size is $\xi_h = -0.14\pm0.17$, consistent with zero.

These values are obtained using sizes and stellar masses based on the S\'{e}rsic photometric model, but the SerExp model produces consistent results, as can be seen in \Fref{fig:halocp}. 
The main differences are that the SerExp model produces a steeper halo mass-stellar mass relation (larger value of parameter $\beta_h$), and that the inference on the halo mass-size correlation is broader and closer to zero ($\xi_h=-0.03\pm0.21$).
The broadening of the posterior probability distribution on parameter $\xi_h$ is related to the smaller intrinsic scatter around the stellar mass-size relation measured for the SerExp model ($\sigma_R = 0.13$ vs. $\sigma_R=0.15$ obtained using S\'{e}rsic): a smaller intrinsic scatter means a shorter lever arm in the measurement of the secondary correlation between $\mhalo$ and $\reff$ at fixed $\mstar$, which in turn results in less precision.

Using the photometric measurements obtained by imposing a de Vaucouleurs profile, we obtain an even steeper stellar-to-halo mass relation, $\beta_h=2.04\pm0.15$, and an average halo mass at the pivot stellar mass (parameter $\mu_{h,0}$) $\sim0.1$~dex higher compared to the S\'{e}rsic case. This difference is related to the different stellar mass distributions obtained with the two models: a de Vaucouleurs profile tends to under-estimate the stellar mass of the most massive CMASS galaxies, for which the data favours values of the S\'{e}rsic index $n > 4$. 
This flattening in the slope of the stellar-to-halo mass relation $\beta$ when going from a de Vaucouleurs to a S\'{e}rsic model is a well known effect \citep[see e.g.]{Sha++14b, Sha++18, Kra++18}.

\begin{figure}
\includegraphics[width=\columnwidth]{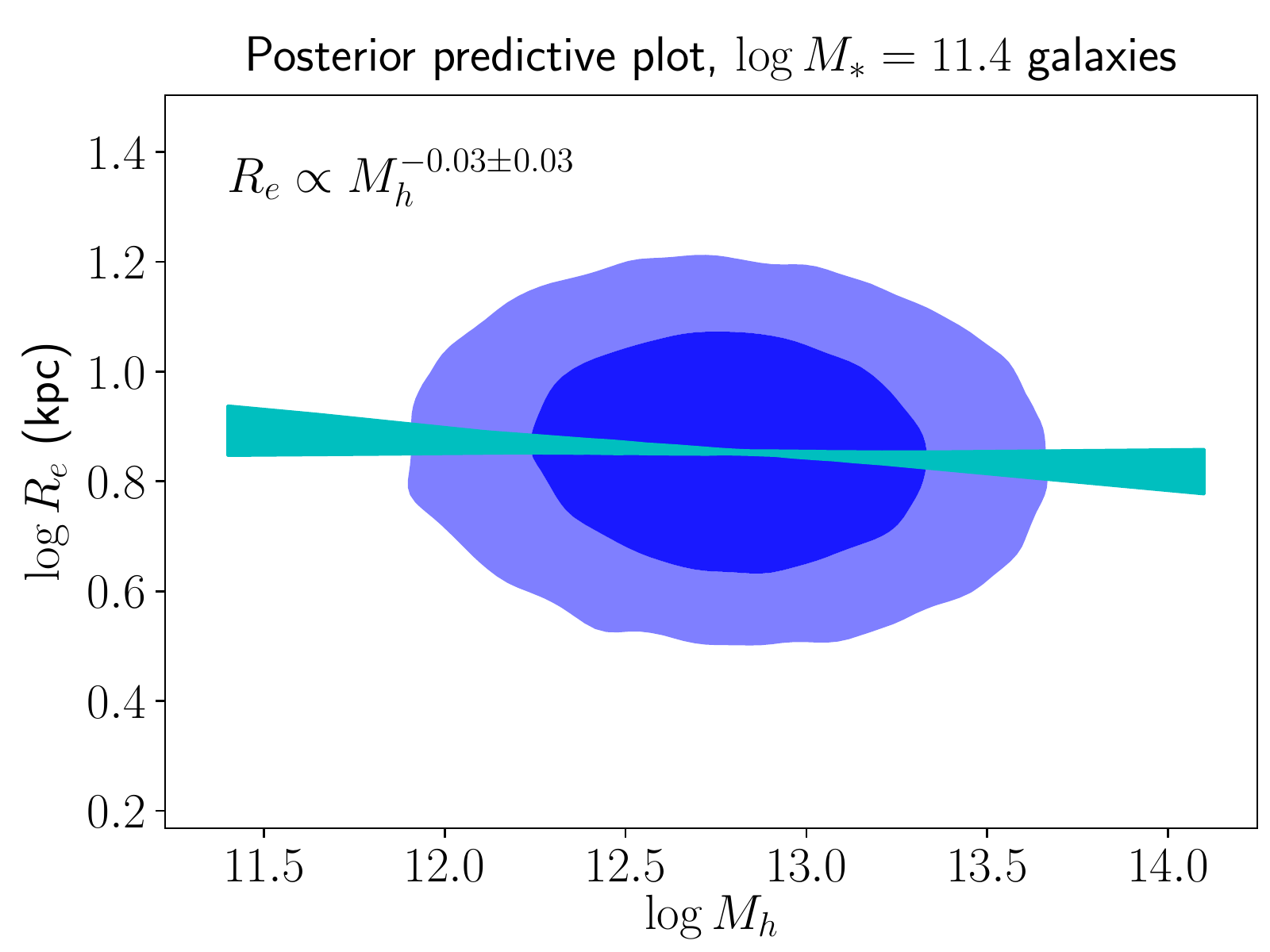}
\caption{
\label{fig:sizem200pp}
Posterior predictive distribution in half-light radius as a function of halo mass for galaxies with $\log{M_*}=11.4$. This is obtained by first drawing values of the hyper-parameters from the posterior probability distribution of our S\'{e}rsic model-based inference, then drawing values of $\reff$ and $\mhalo$ for 1,000 galaxies, given the model specified by the hyper-parameters.
Blue contours correspond to 68\% and 95\% enclosed probability.
The cyan band shows the 68\% credible region of the size-halo mass relation, obtained by fitting a power-law relation to the mock distribution of $\reff$ and $\mhalo$ at each draw of the hyper-parameters. 
}
\end{figure}

In our model of the generalised stellar-to-halo mass relation, stellar mass and size are treated as the independent variables, while halo mass is the dependent variable.
For the purpose of comparing our results with other studies, it can be useful to invert this relation, and obtain the distribution of size as a function of halo mass.
This can be done with a posterior predictive procedure: we fix the value of the stellar mass to $\log{M_*}=11.4$, draw samples of the hyper-parameters from the posterior probability distribution, then, for each value of the hyper-parameters, draw samples of halo mass and half-light radius for the corresponding model.
In \Fref{fig:sizem200pp}, we plot the resulting distribution of $\reff$ as a function of $\mhalo$.
This can be interpreted as our inference of the size-halo mass distribution of $\log{M_*}=11.4$ galaxies, assuming our sample is complete in size at this value of the stellar mass.

We can then summarise this distribution by fitting a power-law halo mass-size relation to it. We do this for each draw of the values of the hyper-parameters, obtaining $\reff \propto \mhalo^{-0.03\pm0.03}$, corresponding to the cyan band in \Fref{fig:sizem200pp}.
This is an alternative visualisation of our main result, consisting of a lack of a strong correlation between halo mass and size at fixed stellar mass.


\subsection{Halo mass as a function of stellar mass, S\'{e}rsic index and size}\label{ssec:halonser}

We now add a dimension to our model: the S\'{e}rsic index.
We modify the galaxy probability distribution \Eref{eq:fulldist} by adding a term describing the distribution in $\nser$, as follows: 
\begin{equation}\label{eq:fullfulldist}
\pr(\mstar,\reff,\mhalo) = \mathcal{S}(\mstar)\mathcal{I}(\nser|\mstar)\mathcal{R}(\reff|\mstar,\nser)\mathcal{H}(\mhalo|\mstar,\nser,\reff).
\end{equation}
The term $\mathcal{I}(\nser|\mstar)$ is the same introduced in \Eref{eq:nserdist}: a Gaussian in the base-10 logarithm of the S\'{e}rsic index, with mean that scales with stellar mass, as parameterised in \Eref{eq:nsermu}.
Then, as done in subsection \ref{ssec:mnr_rel}, we let the mean of the size distribution scale with S\'{e}rsic index, according to \Eref{eq:reffnsermu}.
Finally, we modify the mean of the Gaussian distribution in $\log{\mhalo}$, to allow for a power-law dependence of halo mass on S\'{e}rsic index, as follows:
\begin{equation}\label{eq:muhalonser}
\begin{split}
\mu_h(\mstar,\nser,\reff) = & \mu_{h,0} + \beta_h(\log{\mstar} - 11.4) + \nu_h(\log{\nser} - \mu_n(\mstar)) + \\
& \xi_h(\log{\reff} - \mu_R(\mstar,\nser)).
\end{split}
\end{equation}
In the above equation, $\mu_n(\mstar)$ is the average base-10 logarithm of the S\'{e}rsic index for galaxies of mass $\mstar$, defined in \Eref{eq:nsermu}. The new parameter $\nu_h$, then, describes how halo mass scales with ``excess S\'{e}rsic index'': the ratio between the value of $\nser$ of a galaxy and the typical value of $\nser$ for galaxies of the same stellar mass.

The motivation for allowing halo mass to vary with S\'{e}rsic index is the following: different values of $\nser$ for galaxies of the same stellar mass point to different evolutionary histories, possibly related to the number and type of mergers experienced. If differences in evolutionary paths reflect in the halo mass, we should be able to detect a correlation between $\nser$ and $\mhalo$.

\begin{figure*}
\centering
\includegraphics[width=\textwidth]{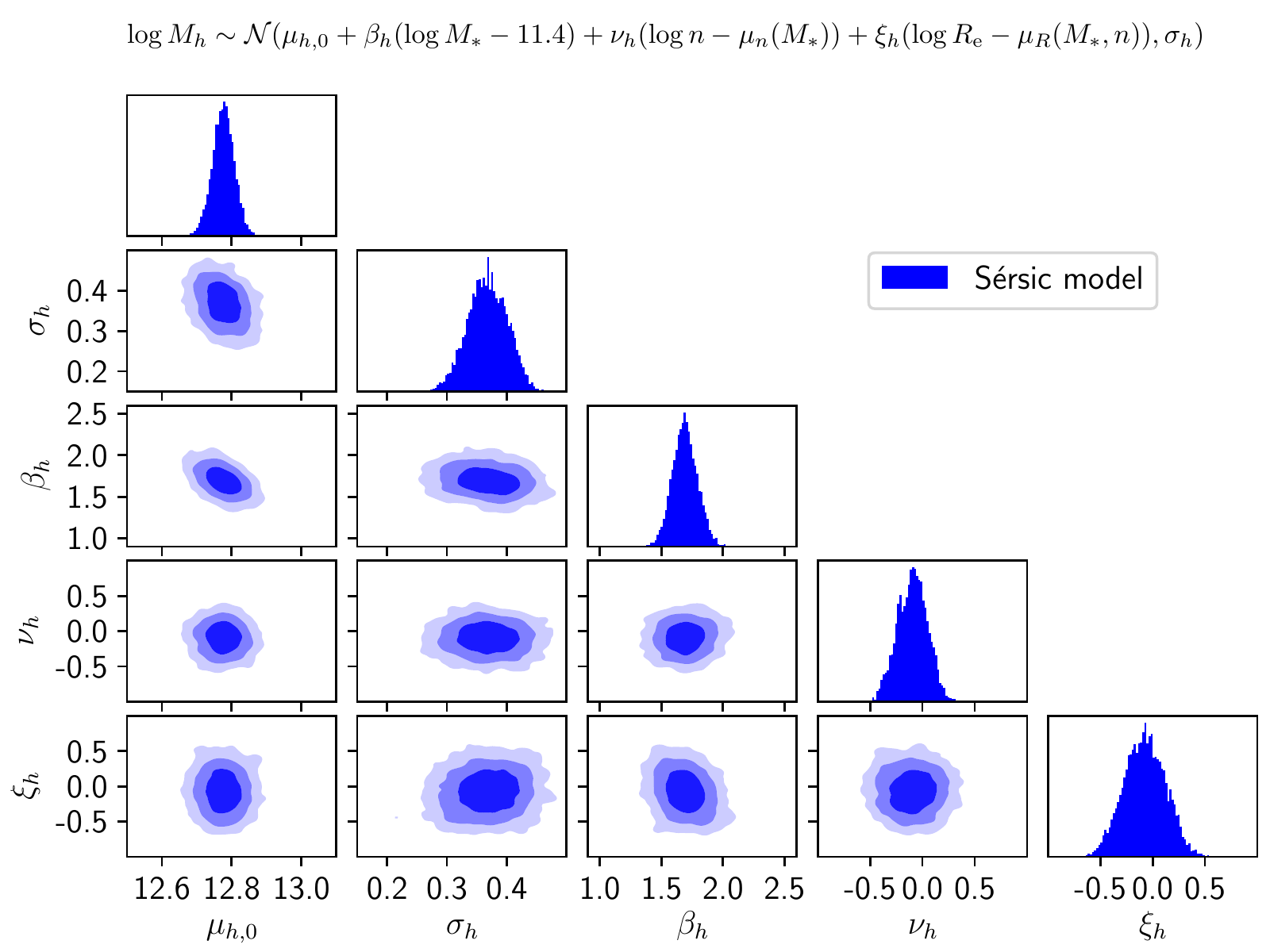}
 \caption{Posterior probability distribution of the parameters describing the distribution of halo mass, for the model including a dependence of $\mhalo$ on the S\'{e}rsic index introduced in subsection \ref{ssec:halonser}. In particular, $\mu_{h,0}$ is the average $\log{\mhalo}$ at a stellar mass $\log{\mstar}=11.4$, average $\nser$ for that mass and average size given $\mstar$ and $\nser$. $\sigma_h$ is the dispersion in $\log{\mhalo}$ around the average, $\beta_h$ is a power-law scaling of halo mass with stellar mass, $\nu_h$ is a power-law scaling of halo mass with excess S\'{e}rsic index (defined as the ratio between the S\'{e}rsic index of a galaxy and the average $\nser$ of galaxies of the same stellar mass) and $\xi_h$ is a power-law scaling of halo mass with excess size.
}
 \label{fig:halonsercp}
\end{figure*}

We run an MCMC to sample the posterior probability distribution of the S\'{e}rsic profile-based model, given the data.
The inference on the hyper-parameters describing the halo mass distribution is shown in \Fref{fig:halonsercp}, while the median values and 68\% credible regions of the inference on all the model hyper-parameters are listed in \Tref{tab:halonser}.

Results change little with respect to the inference based on the simpler model used in the previous subsection.
Most notably, the inference on the new parameter $\nu_h$, describing the correlation between halo mass and S\'{e}rsic index, is consistent with zero: $\nu_h = -0.11\pm0.14$.
Our data, then, allows us to rule out any strong dependence of halo mass on S\'{e}rsic index.
\begin{table*}
\caption{Median values and 68\% credible interval of the posterior probability distribution of the hyper-parameters describing the distribution in stellar mass, half-light radius, S\'{e}rsic index and halo mass of the galaxies in our sample, as inferred by fitting the model introduced in subsection \ref{ssec:halonser} to the S\'{e}rsic profile-based measurements.
}
\label{tab:halonser}
\begin{tabular}{lcl}
\hline
\hline
 & & Parameter description \\
\hline
$\mu_{h,0}$ & $12.78 \pm 0.03$ & Average $\log{M_h}$ at stellar mass $\log{\mstar}=11.4$ and average size\\
$\sigma_h$ & $0.37 \pm 0.03$ & Dispersion in $\log{M_h}$ around the average\\
$\beta_h$ & $1.69 \pm 0.10$ & Power-law dependence of halo mass on $\mstar$ at fixed S\'{e}rsic index and size\\
$\nu_h$ & $-0.11 \pm 0.14$ & Power-law dependence of halo mass on excess S\'{e}rsic index at fixed $M_*$\\
$\xi_h$ & $-0.10 \pm 0.19$ & Power-law dependence of halo mass on excess size at fixed $M_*$ and $\nser$\\
$\mu_{R,0}$ & $0.817 \pm 0.002$ & Average $\log{\reff}$ at stellar mass $\log{\mstar}=11.4$ and S\'{e}rsic index $\nser=4$\\
$\sigma_R$ & $0.128 \pm 0.002$ & Dispersion in $\log{\reff}$ around the average\\
$\beta_R$ & $1.214 \pm 0.012$ & Power-law dependence of size on $\mstar$\\
$\nu_R$ & $0.369 \pm 0.011$ & Power-law dependence of size on $\nser$\\
$\mu_n$ & $0.704 \pm 0.002$ & Average $\log{\nser}$ at stellar mass $\log{\mstar}=11.4$\\
$\sigma_n$ & $0.162 \pm 0.001$ & Dispersion in $\log{\nser}$ around the average\\
$\beta_n$ & $0.467 \pm 0.010$ & Power-law dependence of S\'{e}rsic index on $\mstar$\\
$\mu_*$ & $11.246 \pm 0.004$ & Mean-like parameter of Gaussian term in \Eref{eq:fullskew}\\
$\sigma_*$ & $0.286 \pm 0.004$ & Dispersion-like parameter of Gaussian term in \Eref{eq:fullskew}\\
$\log{s_*}$ & $0.46 \pm 0.02$ & Log of the skewness parameter in \Eref{eq:skew}\\

\hline
\end{tabular}
\end{table*}

\section{Discussion}\label{sect:discuss}

\subsection{The role of observational scatter}\label{ssec:scatter}

The main result of our study is the measurement of the correlation, or lack thereof, between halo mass and galaxy size at fixed stellar mass.
This is a non-trivial measurement to make: galaxies are distributed along a relatively narrow mass-size relation, with a $0.15$~dex intrinsic scatter in $\reff$ at fixed $\mstar$, meaning that any such signal is to be searched across a small dynamic range in size. 
Additionally, observational errors cause data points to move in the $\mstar-\reff$ plane from their true position. If not modelled, these errors can introduce biases: for instance, at fixed observed stellar mass, larger size galaxies are on average more massive than their smaller size counterparts, making the interpretation of correlations measured directly on point estimates of observed quantities problematic (see subsection 2.2 of SL18 for a more detailed explanation). 

The Bayesian hierarchical formalism on which this study is based allows us to forward model the effects of observational scatter, and thus obtain an unbiased measurement of the true distribution of the model parameters, provided that the estimates of the observational uncertainties are accurate.

We obtained uncertainties on the values of $\mstar$ by fitting stellar population synthesis models to HSC photometric data, adding a $0.05$~mag systematic uncertainty to the magnitudes to account for the typical mismatch between the best-fit model and the data, so to avoid unrealistically small errors (see subsection \ref{ssec:ssp}).
As a result, the derived uncertainty on $\mstar$ is to some extent arbitrary, as it reflects our choice on the amplitude of the systematic error added to the data. 
Moreover, the error on $\mstar$ is also set by the priors on the stellar population parameters entering the fit, such as age or dust attenuation.

\begin{figure*}
\centering
\includegraphics[width=\textwidth]{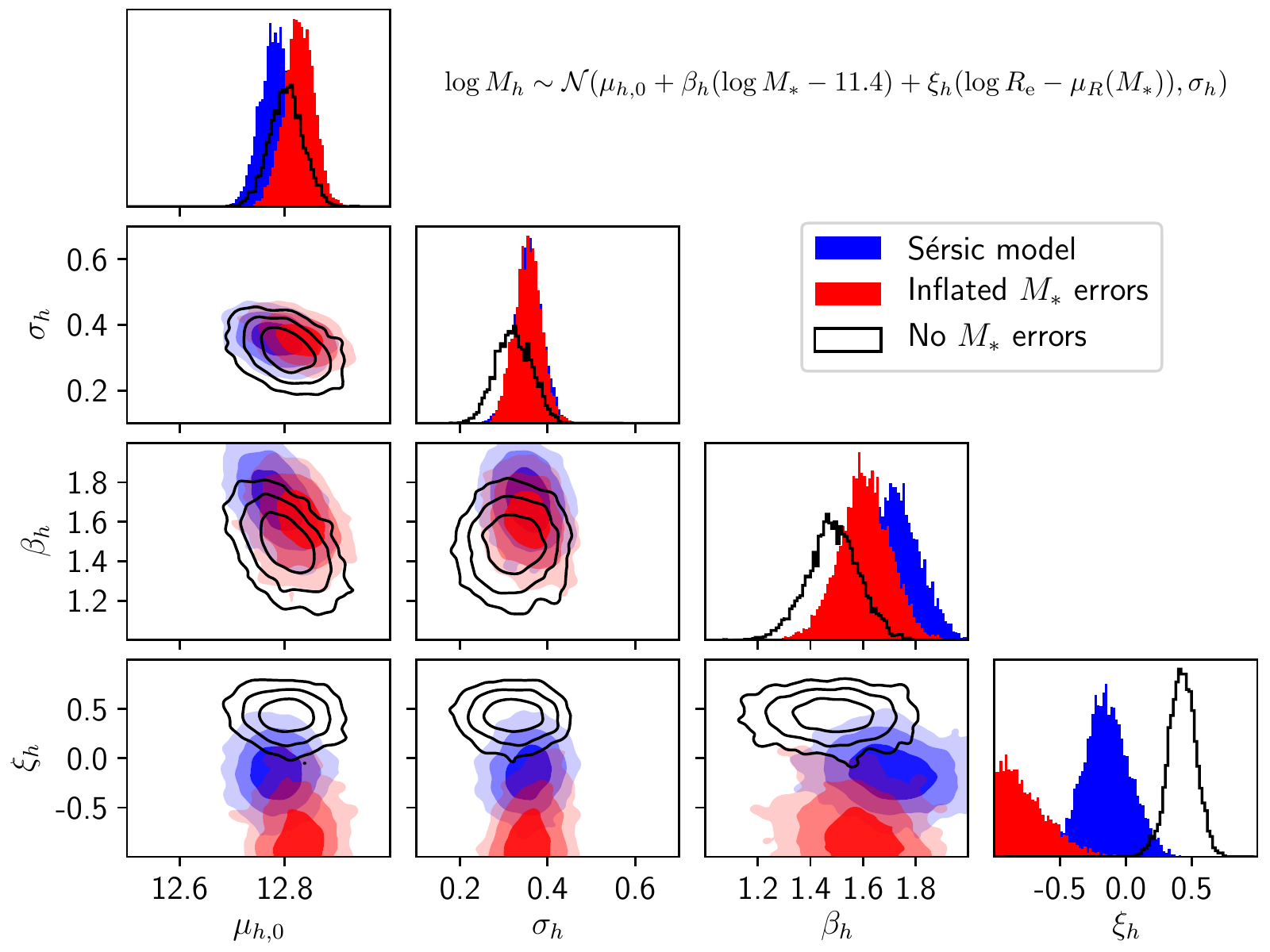}
 \caption{Posterior probability distribution of the parameters describing the distribution of halo mass, based on the model of subsection \ref{ssec:halosize}.
The blue contours show the inference obtained using the S\'{e}rsic model, as obtained in \Sref{sect:halos} and plotted in \Fref{fig:halocp}. The distribution shown in red has been obtained by repeating the analysis of the S\'{e}rsic model and adding $0.1$~dex in quadrature to the uncertainties on the observed stellar masses. The solid lines mark the distribution obtained by setting all uncertainties on the stellar masses to zero.
}
 \label{fig:scattertest}
\end{figure*}

The median value of the uncertainty on $\mstar$ of the sample is $0.10$~dex.
In order to test the impact of making a more conservative choice on the estimate of the stellar mass uncertainty, we repeated the analysis of subsection \ref{ssec:halosize} after adding in quadrature a further $0.10$~dex systematic error on $\mstar$. The inference on the hyper-parameters describing the distribution in halo mass is plotted in red in \Fref{fig:scattertest}, on top of the original inference (in blue). For the sake of simplicity, we only show results based on the S\'{e}rsic model.
The inference on the halo mass-size correlation parameter $\xi_h$ changes: with the larger observational uncertainties on $\mstar$, large negative values of $\xi_h$ are allowed by the data.
However, strong positive correlations between $\mhalo$ and $\reff$ are still ruled out even with inflated error bars on $\mstar$.

For the sake of completeness, we also show the results obtained by setting the uncertainties on $\mstar$ to zero (solid lines in \Fref{fig:scattertest}). In this case, the data seems to favour a positive correlation between halo mass and size. However, as discussed by SL18, this is a biased inference.

\subsection{Comparison with other observational studies}

\subsubsection{The mass-size relation}

A striking result of our study is the steep stellar mass-size relation: as shown in subsection \ref{ssec:mr_rel}, we find $\reff \propto \mstar^{1.37}$ when a single S\'{e}rsic model is used to describe the surface brightness profile of CMASS galaxies ($\reff \propto \mstar^{1.22}$ for the SerExp model).
By comparison, most studies of the mass-size relation in quiescent galaxies find much shallower slopes, with $\beta_R\approx0.5-0.6$ \citep[see e.g.][]{vdW++08, Dam++11, New++12, Hue++13}.
We think the main reason for this discrepancy lies in the stellar mass distribution of our sample. 
The median stellar mass is $\log{\mstar}=11.45$, higher than that of the samples used in the aforementioned studies. As shown by \citet{Ber++11}, the mass-size relation is not a strict power-law at all masses, but becomes steeper above $\mstar\sim2\times10^{11}$, where most of the galaxies in our sample lie.
We are then probing a region of parameter space where the slope of the $\mstar-\reff$ relation is steeper, compared to the value that characterises quiescent galaxies at lower masses.

It is more difficult to reconcile our results with the study of \citet{Fav++18}: in their analysis of the mass-size relation of CMASS galaxies, they find a slope $\beta_R \approx 0.2$, much shallower than our result, and also shallower compared to other studies on different samples of quiescent galaxies, such as the ones mentioned above.
\citet{Fav++18} used de Vaucouleurs profiles to describe early-type galaxies in the CMASS sample: their measurement should then be compared with the value of the mass-size slope we inferred for the same surface brightness profile assumption: $\beta_R=0.98$.

A key difference in the \citet{Fav++18} study is in the stellar mass measurements: while we fitted stellar population synthesis models to HSC photometry, they used stellar masses from the Portsmouth stellar mass catalogue \citep{Mar++13}. The Portsmouth stellar masses are obtained from SDSS photometry, much noisier compared to HSC data. This could lead to a higher observational scatter that would flatten the observed $\mstar-\reff$ relation, if not accounted for.

\subsubsection{The stellar-to-halo mass relation}

\begin{figure}
\includegraphics[width=\columnwidth]{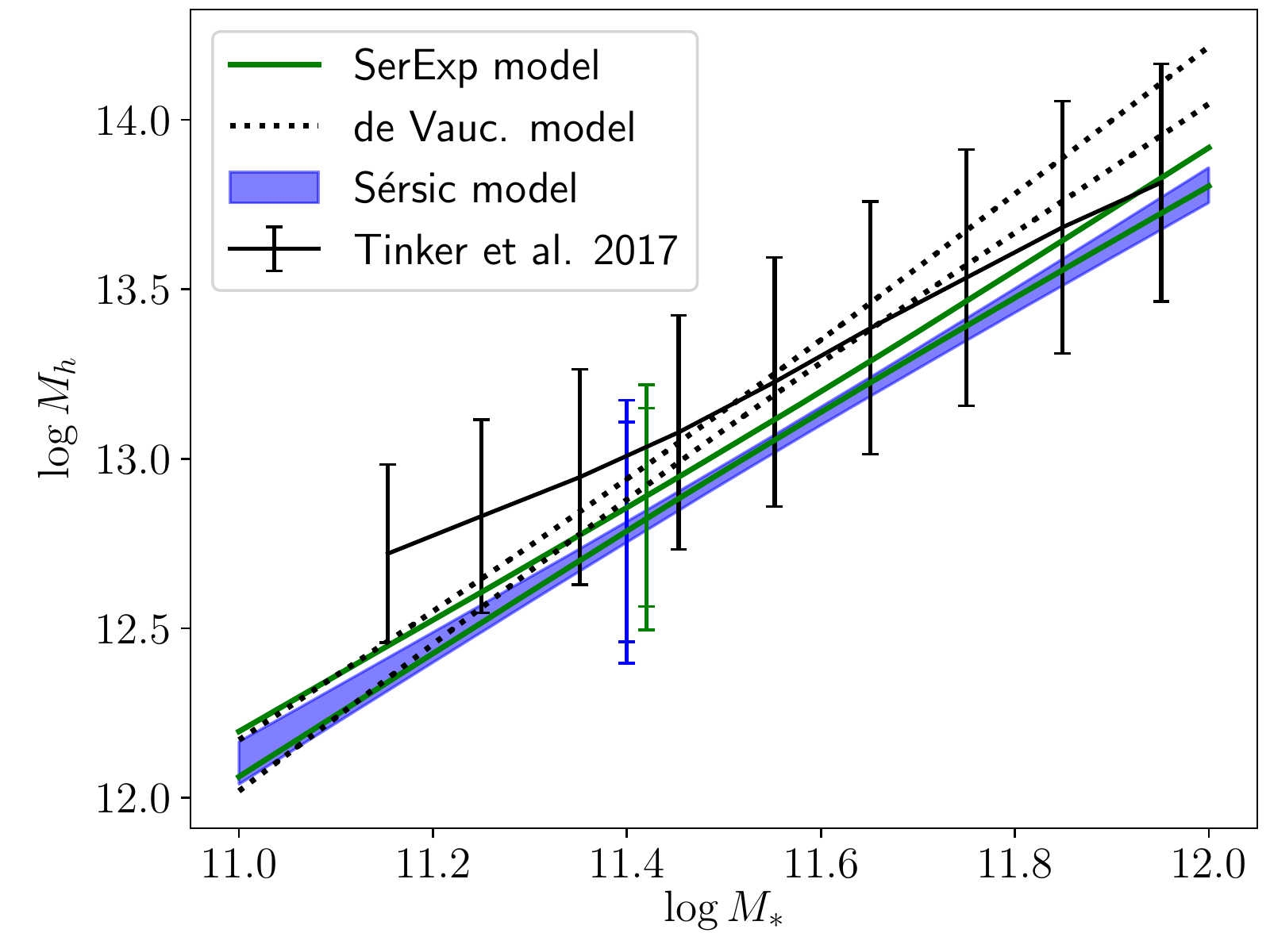}
\caption{Average value of $\log{\mhalo}$ as a function of $\mstar$. The blue band and the green and dotted black pairs of lines show the $1\sigma$ confidence limit obtained from our analysis based on the S\'{e}rsic, SerExp and de Vaucouleurs model, respectively. The solid black line is a measurement from \citet{Tin++17}, obtained from galaxy clustering and abundance matching. Error bars indicate the intrinsic scatter in $\log{\mhalo}$. The inner (outer) ticks on the error bars relative to our measurements refer to the 84 (16) percentile of our inference on the scatter parameter $\sigma_h$. The inference on the intrinsic scatter obtained with the de Vaucouleurs model is omitted, to avoid confusion.}
\label{fig:tinker}
\end{figure}

The stellar-to-halo mass relation of CMASS galaxies was studied by \citet{Tin++17}, using galaxy clustering and abundance matching, and taking advantage of the stellar mass completeness study of \citet{Lea++16}.
In \Fref{fig:tinker} we plot the average value of $\log{\mhalo}$ as a function of $\log{\mstar}$, as measured by \citet{Tin++17}, on top of the same quantity obtained from our inference, calculated using \Eref{eq:muhalo}. 
The stellar masses used by \citet{Tin++17} were obtained under the assumption of a Kroupa IMF, which we converted into Chabrier IMF-based stellar masses by applying a $-0.05$~dex shift.

The clustering-based measurement of \citet{Tin++17} is systematically above our S\'{e}rsic profile-based results by $\sim0.2-0.3$~dex.
This discrepancy is similar to the one observed by \citet{Lea++17}, in their comparison between the clustering and stacked weak lensing signal of the CMASS sample. 
\citet{Lea++17} discussed various possible ways to solve this tension, including varying the value of the cosmological parameter $S_8=\sigma_8\sqrt{\Omega_m/0.3}$, allowing for baryonic physics effects or assembly bias in the models used to map the clustering to the weak lensing signal, allowing for the presence of massive neutrinos or deviations from general relativity.
Part of the tension could also be due to the different photometric data and surface brightness profile used for the stellar mass measurements on which the \citet{Tin++17} study is based: they used cmodel magnitudes obtained from SDSS, while we are showing results based on S\'{e}rsic and SerExp fits on much deeper HSC data. As \Fref{fig:tinker} shows, using a more restrictive de Vaucouleurs profile, which is typically the dominant component in cmodel fits of massive quiescent galaxies, leads to a much better agreement in the average halo mass.

Given all these potential systematics affecting the comparison between clustering and lensing, the discrepancy between our measurement and the \citet{Tin++17} study is not particularly worrisome.
We point out how there is very good agreement on the inference on both the slope of the stellar-to-halo mass relation and the intrinsic scatter, between the two measurements.

\subsubsection{Correlation between galaxy size and halo mass}

Many studies have looked into the correlation between galaxy sizes and the environment they live in, at fixed stellar mass, with somewhat conflicting results (see \Sref{sect:intro}).
Among these studies, the work by \citet{Hua++18} is particularly relevant for our analysis, since it is based on a sample of massive galaxies with photometric data from the HSC survey, selected for having a stellar mass enclosed within a radius of 100~kpc larger than $10^{11.6}M_\odot$ and a spectroscopic redshift in the range $0.3 < z < 0.5$. A good fraction of the galaxies in the \citet{Hua++18} sample also belong to our sample of CMASS galaxies.

\citet{Hua++18} showed how, at fixed value of the stellar mass within 100~kpc in projection, $\mhukpc$, galaxies that are identified as the central of a massive cluster ($\log{\mhalo}\gtrsim 14$) have preferentially smaller values of the stellar mass enclosed within 10~kpc, $\mtenkpc$.
Using $\mhukpc$ as the fiducial value of the stellar mass of a galaxy, this result can be interpreted as the evidence for more extended galaxies (smaller value of $\mtenkpc$ at fixed $\mhukpc$) living preferentially in more massive halos.
A recent stacked weak lensing analysis on the same set of objects confirmed the result \citep{Hua++18b}.

We wish to test whether this trend can be seen in our data as well. 
For this purpose, we perform a posterior predictive test: we use our model to generate mock data, and check whether this mock data reproduces the trend observed by \citet{Hua++18}.
We proceed as follows: we take the maximum-likelihood values of the hyper-parameters describing the distribution in stellar mass, S\'{e}rsic index and size, as measured in subsection \ref{ssec:mnr_rel}, as well as the hyper-parameters describing the distribution in halo mass, as inferred in \Sref{sect:halos} for the S\'{e}rsic case.
For the sake of a more straightforward interpretation of the results, we set the value of the correlation between halo mass and size to zero, $\xi_h=0$, which is consistent with our inference.
We draw a large set of values of $\mstar$, $\nser$, $\reff$, $\mhalo$, then compute the corresponding values of $\mhukpc$ and $\mtenkpc$ for each object.

Both our predicted $\mhukpc$ and $\mstar$ are model-dependent quantities: we are only using data within $8.4''$ to constrain the surface brightness profile of our galaxies, corresponding to a projected physical aperture of $54$~kpc at a redshift $z~0.55$. We are then extrapolating the S\'{e}rsic profile to obtain values of $\mhukpc$ and $\mstar$.

\begin{figure}
\includegraphics[width=\columnwidth]{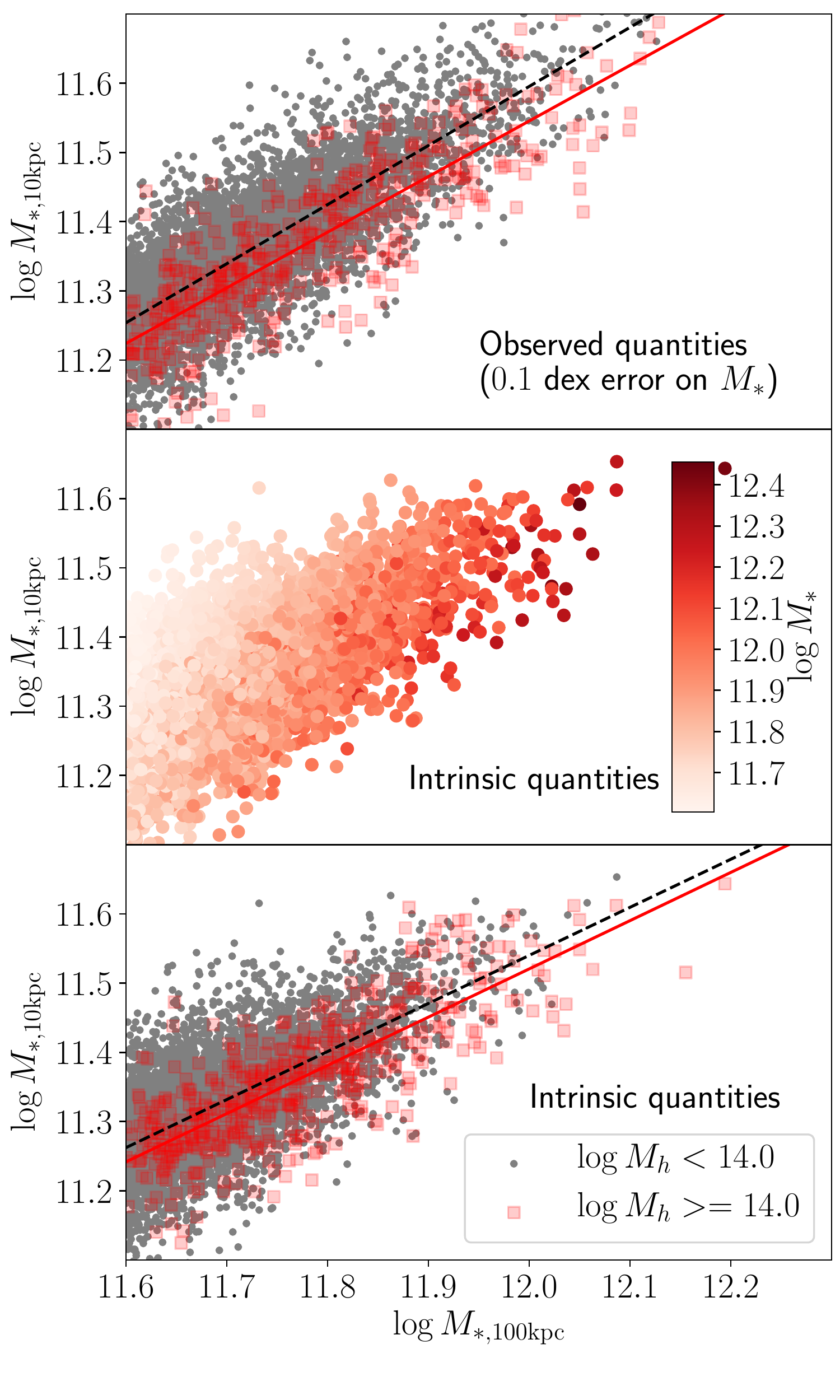}
\caption{Posterior predictive plot, showing the stellar mass enclosed within $10$~kpc as a function of the stellar mass enclosed within $100$~kpc of a mock sample of galaxies, generated from the maximum-likelihood values of the hyper-parameters of our single S\'{e}rsic profile-based model. {\em Bottom:} the sample is split between galaxies that live in halos more massive than $10^{14}M_\odot$ (red squares) and those that live in less massive halos (grey dots). The red and black lines are the best-fit linear relations between $\mtenkpc$ and $\mhukpc$ for the more massive halos and less massive halos respectively.
{\em Middle:} data points from the bottom panel are colour-coded according to the total stellar mass.
{\em Top:} same as the bottom panel, but with a $0.1$~dex random error on the stellar mass measurements.
}
\label{fig:huang}
\end{figure}

In the bottom panel of \Fref{fig:huang}, we plot the values of $\mtenkpc$ as a function of $\mhukpc$. As done in Figure 9 of \citet{Hua++18}, we select objects with $\log{\mhukpc} > 11.6$, split the sample between galaxies that lie in halos more massive than $\log{\mhalo}=14$ (in red) and less massive (in grey), then fit the $\mtenkpc-\mhukpc$ distribution of each subsample with a linear relation.

Qualitatively, we see a similar trend to that reported by \citet{Hua++18}: at fixed $\mhukpc$, galaxies at the centre of massive halos have preferentially lower values of $\mtenkpc$.
This might seem in contradiction with the way our model was built, since we removed any dependence of the halo mass on size at fixed stellar mass.
However, in the context of the S\'{e}rsic models at the basis of our analysis, a non-negligible fraction of the total stellar mass of a galaxy comes from stars located beyond $100$~kpc in projection, especially for very massive galaxies like the ones considered in this experiment. 
The fraction of stellar mass beyond $100$~kpc is larger for galaxies with larger sizes: this implies that, at fixed $\mhukpc$, galaxies with larger sizes, or with smaller values of $\mtenkpc$, are on average more massive (have a larger value of $\mstar$). As a result, these galaxies tend to live in more massive dark matter halos, on average.
This can verified in the middle panel of \Fref{fig:huang}, where the same data points of the plot in the bottom panel are colour-coded by the total stellar mass of each object: stellar mass increases with increasing $\mhukpc$, but also with decreasing $\mtenkpc$. Halo mass follows a similar trend.

In the top panel of \Fref{fig:huang}, we show a version of the bottom plot of the same figure, modified by adding a $0.1$~dex random observational scatter on the stellar mass measurements. This observational scatter is meant to simulate errors in the stellar population synthesis fitting, therefore they affect the measurements of the stellar mass at different radii in the same way (i.e. they shift the values of $\log{\mhukpc}$ and $\log{\mtenkpc}$ by the exactly same amount).
The difference in the $\mtenkpc-\mhukpc$ relation between clusters and lower-mass halos increases when looking at observed quantities, and is similar in amplitude to the signal measured by \citet{Hua++18}. 
This is related to the distortion of the mass-size relation due to observational scatter discussed in subsection \ref{ssec:scatter} and can be understood as follows: given a bin in observed stellar mass, smaller sized galaxies are statistically more likely to have scattered into the bin from lower intrinsic stellar masses, and therefore to live in less massive dark matter halos, compared to larger sized ones (see also subsection 2.2 of SL18).

Both of the effects discussed above follow simply from the existence of a positive correlation between stellar mass and size and do not depend on a particular form of the mass-size or mass-S\'{e}rsic index relation. Indeed, we have verified that making the same posterior predictive plot based on de Vaucouleurs fits produces very similar results.

After this test, we conclude that our inference is consistent with the \citet{Hua++18} analysis. In the context of our model, the correlation between halo mass and the observed values of $\mtenkpc$ at fixed $\mhukpc$ can be explained with the combination of two causes: massive galaxies having a non-negligible fraction of their stellar mass beyond $100$~kpc, and the effect of observational scatter on the stellar mass measurements.

\citet{Cha++17} measured the correlation between halo mass and galaxy size at fixed stellar mass, using stacked weak lensing measurements around a large set of galaxies. \citet{Cha++17} made bins in luminosity and size, and looked at the variation in halo mass, as measured from weak lensing, as a function of size, after accounting for the dependence of halo mass on stellar mass.
They measured a value $\xi_h=0.42\pm0.12$, inconsistent with our inference.
SL18 argued that part of their signal could be due to the effects of observational scatter. This would definitely be the case if their stacked weak lensing analysis was carried out in bins of observed stellar mass: as \Fref{fig:scattertest} shows, ignoring observational errors on $\mstar$ can introduce a signal of similar magnitude to the reported value. However, \citet{Cha++17} carry our their analysis in luminosity bins. Luminosity is measured with a much greater accuracy than stellar mass, and this should lead to a more accurate inference on the parameter $\xi_h$. Nevertheless, the \citet{Cha++17} analysis still relies on noisy stellar mass measurements from stellar population synthesis modelling, and that should have some impact on the inference, if not taken into account. 
For a fair comparison, it would be important to test the effects of observational scatter in a study of the halo mass-size correlation \`{a} la \citet{Cha++17}. This, however, is beyond the scope of this paper.

\subsection{Comparison with theoretical models}

The relation between halo mass and galaxy size has been investigated in theoretical works based on both semi-analytical models or hydrodynamical simulations.
\citet{Sha++14} explored various different semi-analytical models, finding that most of these models predict a positive correlation between halo mass and size at fixed stellar mass. 
They showed that such a correlation can emerge simply as the result of dissipationless mergers: although the major merger rate is a very mild function of halo mass, the rate of minor mergers can be higher in higher mass halos, provided that the merging timescale, which is set by dynamical friction, is sufficiently short.
For a given increase in stellar mass, minor mergers produce a higher increase in size compared to major mergers \citep{Naa++09, Hop++10}: if minor mergers are more frequent in more massive halos, then this can introduce a halo mass-size correlation at fixed stellar mass. 
Whether this effect is real, however, depends on the merging timescale and its dependence on the mass of the central halo and that of the accreted object, which is very difficult to determine observationally \citep[see also][for related discussions on merger time scales and the effects of mergers on the size evolution of massive galaxies]{Hop++10b, Nip++12}.
Assuming dynamical friction timescales from \citet{McC++12}, the semi-empirical hierarchical growth model of \citet{Sha++15} found only a mild dependence on halo mass in the size growth of massive galaxies.

Other processes that can produce a positive correlation between halo mass and size at fixed stellar mass are disc instabilities, gas dissipation in major mergers and the differential evolution of accreted galaxies in low and high mass halos. We refer to \citet{Sha++14} for details on these mechanisms.
\citet{Sha++14} claimed that models predicting a significant correlation between halo mass and galaxy size are disfavoured by observational data, based on evidence from the work by \citet{Hue++13b}. Our analysis further strengthens their point.

We also point out that, although we are focusing our comparison on theoretical models based on a merger-driven size growth scenario, it is possible that the most massive galaxies observed at $z<1$ have assembled most of their mass through in-situ processes during their early evolutionary phases \citep[see][for a related discussion]{B+S16}. In that case, it is not clear if a correlation between size (or S\'{e}rsic index) and halo mass at fixed stellar mass is expected at all.

\citet{Cha++17} looked at the halo mass-size correlation at fixed $\mstar$ in the EAGLE \citep{Sch++15} and ILLUSTRIS \citep{Vog++14}. They found values of $\xi_h$ in the range $0.5 \lesssim \xi_h \lesssim 1.0$ at the stellar masses probed by our study, although they find smaller values when restricting their measurement to central galaxies. For instance, they report a value of $\xi_h = 0.29\pm0.23$ for EAGLE galaxies in a bin of median stellar mass $\log{\mstar} = 11.34$, once satellites are excluded. A similar value ($\xi_h = 0.18\pm0.07$) was measured by \citet{Des++17} on the same simulation.
Our measurements are in agreement with these values to within $2\sigma$, therefore we do not register any serious tension between predictions from hydrodynamical simulations and our data. 


\section{Conclusions}\label{sect:concl}

We used HSC imaging and weak lensing measurements for a set of $\sim10,000$ galaxies from the CMASS sample to constrain 1) the stellar mass-size relation, 2) the stellar mass-S\'{e}rsic index-size relation, 3) the stellar mass-size-halo mass relation and 4) the stellar mass-size-S\'{e}rsic index-halo mass relation of massive ($\log{\mstar} > 11$) galaxies at redshift $z\sim0.55$.
We found the following:
\begin{itemize}
\item The stellar mass-size relation at the high mass end is very steep: modelling this as a power-law, $\mstar \propto R^\beta_R$, we find values of $\beta_R$ larger than unity, with the actual value depending on the choice of surface brightness profile used to fit the photometry.
\item The average S\'{e}rsic index of our sample increases with increasing stellar mass as $\nser \propto \mstar^{0.46}$. At fixed stellar mass, size correlates positively with S\'{e}rsic index.
\item Halo mass scales with the $\sim1.7$ power of stellar mass. At fixed stellar mass, we find no evidence for an additional correlation of halo mass with size or S\'{e}rsic index. As a consequence, our measurement disfavours models of galaxy evolution in which galaxy size grows more efficiently in high mass halos compared to low mass halos.
\item The halo masses of the CMASS sample inferred from weak lensing are on average $0.2-0.3$~dex smaller than those inferred from galaxy clustering, though much of this discrepancy could be due to differences in the photometric data and the choice of the surface brightness profile used to model galaxies, on which stellar mass measurements are based.
\end{itemize}

The observational data at the core of this work is the weak lensing measurements from HSC. This data consists of observations carried out in the first year of the survey, covering $\sim140$~deg$^2$. As the HSC survey reaches its completion, similar quality measurements over a much larger area will soon be available. This should allow us to reduce statistical uncertainties on the distribution of halo masses.
However, for the key parameter subject of this study, the correlation between halo mass and size at fixed stellar mass $\xi_h$, the inferred value depends in a non-negligible way on the surface brightness profile used to fit the photometric data, on which the stellar mass and size measurements are based. 
As statistical errors shrink, the systematic effect connected to this aspect of the model will start to dominate. 
We point out that, as shown in subsection \ref{ssec:serexp}, most of the difference between the measurements based on the S\'{e}rsic and SerExp models is due to the extrapolation of the surface brightness profile at large radii, where the model is unconstrained by the data.
We then expect that the next challenge in the determination of correlations between the stellar distribution in galaxies and their host dark matter halos will be the measurement of the light profile in the very outskirts of galaxies.

\begin{acknowledgements}

AS acknowledges funding from the European Union's Horizon 2020 research and innovation programme under grant agreement No 792916, as well as a KAKENHI Grant from the Japan Society for the Promotion of Science, MEXT, No JP17K14250.
This work was supported by World Premier International Research Center Initiative (WPI Initiative), MEXT, Japan.

The Hyper Suprime-Cam (HSC) collaboration includes the astronomical communities of Japan and Taiwan, and Princeton University.  The HSC instrumentation and software were developed by the National Astronomical Observatory of Japan (NAOJ), the Kavli Institute for the Physics and Mathematics of the Universe (Kavli IPMU), the University of Tokyo, the High Energy Accelerator Research Organization (KEK), the Academia Sinica Institute for Astronomy and Astrophysics in Taiwan (ASIAA), and Princeton University.  Funding was contributed by the FIRST program from Japanese Cabinet Office, the Ministry of Education, Culture, Sports, Science and Technology (MEXT), the Japan Society for the Promotion of Science (JSPS),  Japan Science and Technology Agency  (JST),  the Toray Science  Foundation, NAOJ, Kavli IPMU, KEK, ASIAA,  and Princeton University.

Funding for SDSS-III has been provided by the Alfred P. Sloan Foundation, the Participating Institutions, the National Science Foundation, and the U.S. Department of Energy Office of Science. The SDSS-III web site is http://www.sdss3.org/.

SDSS-III is managed by the Astrophysical Research Consortium for the Participating Institutions of the SDSS-III Collaboration including the University of Arizona, the Brazilian Participation Group, Brookhaven National Laboratory, Carnegie Mellon University, University of Florida, the French Participation Group, the German Participation Group, Harvard University, the Instituto de Astrofisica de Canarias, the Michigan State/Notre Dame/JINA Participation Group, Johns Hopkins University, Lawrence Berkeley National Laboratory, Max Planck Institute for Astrophysics, Max Planck Institute for Extraterrestrial Physics, New Mexico State University, New York University, Ohio State University, Pennsylvania State University, University of Portsmouth, Princeton University, the Spanish Participation Group, University of Tokyo, University of Utah, Vanderbilt University, University of Virginia, University of Washington, and Yale University.

\end{acknowledgements}

\bibliographystyle{aa}
\bibliography{references}

\begin{thebibliography}{77}
\expandafter\ifx\csname natexlab\endcsname\relax\def\natexlab#1{#1}\fi

\bibitem[{{Aihara} {et~al.}(2018{\natexlab{a}}){Aihara}, {Arimoto},
  {Armstrong}, {Arnouts}, {Bahcall}, {Bickerton}, {Bosch}, {Bundy}, {Capak},
  {Chan}, {Chiba}, {Coupon}, {Egami}, {Enoki}, {Finet}, {Fujimori}, {Fujimoto},
  {Furusawa}, {Furusawa}, {Goto}, {Goulding}, {Greco}, {Greene}, {Gunn},
  {Hamana}, {Harikane}, {Hashimoto}, {Hattori}, {Hayashi}, {Hayashi},
  {He{\l}miniak}, {Higuchi}, {Hikage}, {Ho}, {Hsieh}, {Huang}, {Huang},
  {Ikeda}, {Imanishi}, {Inoue}, {Iwasawa}, {Iwata}, {Jaelani}, {Jian},
  {Kamata}, {Karoji}, {Kashikawa}, {Katayama}, {Kawanomoto}, {Kayo}, {Koda},
  {Koike}, {Kojima}, {Komiyama}, {Konno}, {Koshida}, {Koyama}, {Kusakabe},
  {Leauthaud}, {Lee}, {Lin}, {Lin}, {Lupton}, {Mandelbaum}, {Matsuoka},
  {Medezinski}, {Mineo}, {Miyama}, {Miyatake}, {Miyazaki}, {Momose}, {More},
  {More}, {Moritani}, {Moriya}, {Morokuma}, {Mukae}, {Murata}, {Murayama},
  {Nagao}, {Nakata}, {Niida}, {Niikura}, {Nishizawa}, {Obuchi}, {Oguri},
  {Oishi}, {Okabe}, {Okamoto}, {Okura}, {Ono}, {Onodera}, {Onoue}, {Osato},
  {Ouchi}, {Price}, {Pyo}, {Sako}, {Sawicki}, {Shibuya}, {Shimasaku},
  {Shimono}, {Shirasaki}, {Silverman}, {Simet}, {Speagle}, {Spergel},
  {Strauss}, {Sugahara}, {Sugiyama}, {Suto}, {Suyu}, {Suzuki}, {Tait},
  {Takada}, {Takata}, {Tamura}, {Tanaka}, {Tanaka}, {Tanaka}, {Tanaka},
  {Terai}, {Terashima}, {Toba}, {Tominaga}, {Toshikawa}, {Turner}, {Uchida},
  {Uchiyama}, {Umetsu}, {Uraguchi}, {Urata}, {Usuda}, {Utsumi}, {Wang}, {Wang},
  {Wong}, {Yabe}, {Yamada}, {Yamanoi}, {Yasuda}, {Yeh}, {Yonehara}, \&
  {Yuma}}]{Aih++18a}
{Aihara}, H., {Arimoto}, N., {Armstrong}, R., {et~al.} 2018{\natexlab{a}},
  \pasj, 70, S4

\bibitem[{{Aihara} {et~al.}(2018{\natexlab{b}}){Aihara}, {Armstrong},
  {Bickerton}, {Bosch}, {Coupon}, {Furusawa}, {Hayashi}, {Ikeda}, {Kamata},
  {Karoji}, {Kawanomoto}, {Koike}, {Komiyama}, {Lang}, {Lupton}, {Mineo},
  {Miyatake}, {Miyazaki}, {Morokuma}, {Obuchi}, {Oishi}, {Okura}, {Price},
  {Takata}, {Tanaka}, {Tanaka}, {Tanaka}, {Uchida}, {Uraguchi}, {Utsumi},
  {Wang}, {Yamada}, {Yamanoi}, {Yasuda}, {Arimoto}, {Chiba}, {Finet},
  {Fujimori}, {Fujimoto}, {Furusawa}, {Goto}, {Goulding}, {Gunn}, {Harikane},
  {Hattori}, {Hayashi}, {He{\l}miniak}, {Higuchi}, {Hikage}, {Ho}, {Hsieh},
  {Huang}, {Huang}, {Imanishi}, {Iwata}, {Jaelani}, {Jian}, {Kashikawa},
  {Katayama}, {Kojima}, {Konno}, {Koshida}, {Kusakabe}, {Leauthaud}, {Lee},
  {Lin}, {Lin}, {Mandelbaum}, {Matsuoka}, {Medezinski}, {Miyama}, {Momose},
  {More}, {More}, {Mukae}, {Murata}, {Murayama}, {Nagao}, {Nakata}, {Niida},
  {Niikura}, {Nishizawa}, {Oguri}, {Okabe}, {Ono}, {Onodera}, {Onoue}, {Ouchi},
  {Pyo}, {Shibuya}, {Shimasaku}, {Simet}, {Speagle}, {Spergel}, {Strauss},
  {Sugahara}, {Sugiyama}, {Suto}, {Suzuki}, {Tait}, {Takada}, {Terai}, {Toba},
  {Turner}, {Uchiyama}, {Umetsu}, {Urata}, {Usuda}, {Yeh}, \&
  {Yuma}}]{Aih++18b}
{Aihara}, H., {Armstrong}, R., {Bickerton}, S., {et~al.} 2018{\natexlab{b}},
  \pasj, 70, S8

\bibitem[{{Allen} {et~al.}(2015){Allen}, {Kacprzak}, {Spitler}, {Glazebrook},
  {Labb{\'e}}, {Tran}, {Straatman}, {Nanayakkara}, {Brammer}, {Quadri},
  {Cowley}, {Monson}, {Papovich}, {Persson}, {Rees}, {Tilvi}, \&
  {Tomczak}}]{All++15}
{Allen}, R.~J., {Kacprzak}, G.~G., {Spitler}, L.~R., {et~al.} 2015, \apj, 806,
  3

\bibitem[{{Anderson} {et~al.}(2014){Anderson}, {Aubourg}, {Bailey}, {Beutler},
  {Bhardwaj}, {Blanton}, {Bolton}, {Brinkmann}, {Brownstein}, {Burden},
  {Chuang}, {Cuesta}, {Dawson}, {Eisenstein}, {Escoffier}, {Gunn}, {Guo}, {Ho},
  {Honscheid}, {Howlett}, {Kirkby}, {Lupton}, {Manera}, {Maraston}, {McBride},
  {Mena}, {Montesano}, {Nichol}, {Nuza}, {Olmstead}, {Padmanabhan},
  {Palanque-Delabrouille}, {Parejko}, {Percival}, {Petitjean}, {Prada},
  {Price-Whelan}, {Reid}, {Roe}, {Ross}, {Ross}, {Sabiu}, {Saito}, {Samushia},
  {S{\'a}nchez}, {Schlegel}, {Schneider}, {Scoccola}, {Seo}, {Skibba},
  {Strauss}, {Swanson}, {Thomas}, {Tinker}, {Tojeiro}, {Maga{\~n}a}, {Verde},
  {Wake}, {Weaver}, {Weinberg}, {White}, {Xu}, {Y{\`e}che}, {Zehavi}, \&
  {Zhao}}]{And++14}
{Anderson}, L., {Aubourg}, {\'E}., {Bailey}, S., {et~al.} 2014, \mnras, 441, 24

\bibitem[{{Auger} {et~al.}(2009){Auger}, {Treu}, {Bolton}, {Gavazzi},
  {Koopmans}, {Marshall}, {Bundy}, \& {Moustakas}}]{Aug++09}
{Auger}, M.~W., {Treu}, T., {Bolton}, A.~S., {et~al.} 2009, \apj, 705, 1099

\bibitem[{{Axelrod} {et~al.}(2010){Axelrod}, {Kantor}, {Lupton}, \&
  {Pierfederici}}]{Axe++10}
{Axelrod}, T., {Kantor}, J., {Lupton}, R.~H., \& {Pierfederici}, F. 2010, in
  \procspie, Vol. 7740, Software and Cyberinfrastructure for Astronomy, 774015

\bibitem[{{Bernardi} {et~al.}(2014){Bernardi}, {Meert}, {Vikram},
  {Huertas-Company}, {Mei}, {Shankar}, \& {Sheth}}]{Ber++14}
{Bernardi}, M., {Meert}, A., {Vikram}, V., {et~al.} 2014, \mnras, 443, 874

\bibitem[{{Bernardi} {et~al.}(2011){Bernardi}, {Roche}, {Shankar}, \&
  {Sheth}}]{Ber++11}
{Bernardi}, M., {Roche}, N., {Shankar}, F., \& {Sheth}, R.~K. 2011, \mnras,
  412, L6

\bibitem[{{Bertin} \& {Arnouts}(1996)}]{B+A96}
{Bertin}, E. \& {Arnouts}, S. 1996, \aaps, 117, 393

\bibitem[{{Beutler} {et~al.}(2014){Beutler}, {Saito}, {Seo}, {Brinkmann},
  {Dawson}, {Eisenstein}, {Font-Ribera}, {Ho}, {McBride}, {Montesano},
  {Percival}, {Ross}, {Ross}, {Samushia}, {Schlegel}, {S{\'a}nchez}, {Tinker},
  \& {Weaver}}]{Beu++14}
{Beutler}, F., {Saito}, S., {Seo}, H.-J., {et~al.} 2014, \mnras, 443, 1065

\bibitem[{{Bosch} {et~al.}(2018){Bosch}, {Armstrong}, {Bickerton}, {Furusawa},
  {Ikeda}, {Koike}, {Lupton}, {Mineo}, {Price}, {Takata}, {Tanaka}, {Yasuda},
  {AlSayyad}, {Becker}, {Coulton}, {Coupon}, {Garmilla}, {Huang}, {Krughoff},
  {Lang}, {Leauthaud}, {Lim}, {Lust}, {MacArthur}, {Mandelbaum}, {Miyatake},
  {Miyazaki}, {Murata}, {More}, {Okura}, {Owen}, {Swinbank}, {Strauss},
  {Yamada}, \& {Yamanoi}}]{Bos++18}
{Bosch}, J., {Armstrong}, R., {Bickerton}, S., {et~al.} 2018, \pasj, 70, S5

\bibitem[{{Bruzual} \& {Charlot}(2003)}]{B+C03}
{Bruzual}, G. \& {Charlot}, S. 2003, \mnras, 344, 1000

\bibitem[{{Buchan} \& {Shankar}(2016)}]{B+S16}
{Buchan}, S. \& {Shankar}, F. 2016, \mnras, 462, 2001

\bibitem[{{Caon} {et~al.}(1993){Caon}, {Capaccioli}, \& {D'Onofrio}}]{CCD93}
{Caon}, N., {Capaccioli}, M., \& {D'Onofrio}, M. 1993, \mnras, 265, 1013

\bibitem[{{Chabrier}(2003)}]{Cha03}
{Chabrier}, G. 2003, \pasp, 115, 763

\bibitem[{{Charlton} {et~al.}(2017){Charlton}, {Hudson}, {Balogh}, \&
  {Khatri}}]{Cha++17}
{Charlton}, P.~J.~L., {Hudson}, M.~J., {Balogh}, M.~L., \& {Khatri}, S. 2017,
  \mnras, 472, 2367

\bibitem[{{Ciotti} \& {Bertin}(1999)}]{C+B99}
{Ciotti}, L. \& {Bertin}, G. 1999, \aap, 352, 447

\bibitem[{{Cooper} {et~al.}(2012){Cooper}, {Griffith}, {Newman}, {Coil},
  {Davis}, {Dutton}, {Faber}, {Guhathakurta}, {Koo}, {Lotz}, {Weiner},
  {Willmer}, \& {Yan}}]{Coo++12}
{Cooper}, M.~C., {Griffith}, R.~L., {Newman}, J.~A., {et~al.} 2012, \mnras,
  419, 3018

\bibitem[{{Daddi} {et~al.}(2005){Daddi}, {Renzini}, {Pirzkal}, {Cimatti},
  {Malhotra}, {Stiavelli}, {Xu}, {Pasquali}, {Rhoads}, {Brusa}, {di Serego
  Alighieri}, {Ferguson}, {Koekemoer}, {Moustakas}, {Panagia}, \&
  {Windhorst}}]{Dad++05}
{Daddi}, E., {Renzini}, A., {Pirzkal}, N., {et~al.} 2005, \apj, 626, 680

\bibitem[{{Damjanov} {et~al.}(2011){Damjanov}, {Abraham}, {Glazebrook},
  {McCarthy}, {Caris}, {Carlberg}, {Chen}, {Crampton}, {Green}, {J{\o}rgensen},
  {Juneau}, {Le Borgne}, {Marzke}, {Mentuch}, {Murowinski}, {Roth}, {Savaglio},
  \& {Yan}}]{Dam++11}
{Damjanov}, I., {Abraham}, R.~G., {Glazebrook}, K., {et~al.} 2011, \apjl, 739,
  L44

\bibitem[{{Damjanov} {et~al.}(2015){Damjanov}, {Zahid}, {Geller}, \&
  {Hwang}}]{Dam++15}
{Damjanov}, I., {Zahid}, H.~J., {Geller}, M.~J., \& {Hwang}, H.~S. 2015, \apj,
  815, 104

\bibitem[{{Dawson} {et~al.}(2013){Dawson}, {Schlegel}, {Ahn}, {Anderson},
  {Aubourg}, {Bailey}, {Barkhouser}, {Bautista}, {Beifiori}, {Berlind},
  {Bhardwaj}, {Bizyaev}, {Blake}, {Blanton}, {Blomqvist}, {Bolton}, {Borde},
  {Bovy}, {Brandt}, {Brewington}, {Brinkmann}, {Brown}, {Brownstein}, {Bundy},
  {Busca}, {Carithers}, {Carnero}, {Carr}, {Chen}, {Comparat}, {Connolly},
  {Cope}, {Croft}, {Cuesta}, {da Costa}, {Davenport}, {Delubac}, {de Putter},
  {Dhital}, {Ealet}, {Ebelke}, {Eisenstein}, {Escoffier}, {Fan}, {Filiz Ak},
  {Finley}, {Font-Ribera}, {G{\'e}nova-Santos}, {Gunn}, {Guo}, {Haggard},
  {Hall}, {Hamilton}, {Harris}, {Harris}, {Ho}, {Hogg}, {Holder}, {Honscheid},
  {Huehnerhoff}, {Jordan}, {Jordan}, {Kauffmann}, {Kazin}, {Kirkby}, {Klaene},
  {Kneib}, {Le Goff}, {Lee}, {Long}, {Loomis}, {Lundgren}, {Lupton}, {Maia},
  {Makler}, {Malanushenko}, {Malanushenko}, {Mandelbaum}, {Manera}, {Maraston},
  {Margala}, {Masters}, {McBride}, {McDonald}, {McGreer}, {McMahon}, {Mena},
  {Miralda-Escud{\'e}}, {Montero-Dorta}, {Montesano}, {Muna}, {Myers},
  {Naugle}, {Nichol}, {Noterdaeme}, {Nuza}, {Olmstead}, {Oravetz}, {Oravetz},
  {Owen}, {Padmanabhan}, {Palanque-Delabrouille}, {Pan}, {Parejko},
  {P{\^a}ris}, {Percival}, {P{\'e}rez-Fournon}, {P{\'e}rez-R{\`a}fols},
  {Petitjean}, {Pfaffenberger}, {Pforr}, {Pieri}, {Prada}, {Price-Whelan},
  {Raddick}, {Rebolo}, {Rich}, {Richards}, {Rockosi}, {Roe}, {Ross}, {Ross},
  {Rossi}, {Rubi{\~n}o-Martin}, {Samushia}, {S{\'a}nchez}, {Sayres}, {Schmidt},
  {Schneider}, {Sc{\'o}ccola}, {Seo}, {Shelden}, {Sheldon}, {Shen}, {Shu},
  {Slosar}, {Smee}, {Snedden}, {Stauffer}, {Steele}, {Strauss}, {Streblyanska},
  {Suzuki}, {Swanson}, {Tal}, {Tanaka}, {Thomas}, {Tinker}, {Tojeiro},
  {Tremonti}, {Vargas Maga{\~n}a}, {Verde}, {Viel}, {Wake}, {Watson}, {Weaver},
  {Weinberg}, {Weiner}, {West}, {White}, {Wood-Vasey}, {Yeche}, {Zehavi},
  {Zhao}, \& {Zheng}}]{Daw++13}
{Dawson}, K.~S., {Schlegel}, D.~J., {Ahn}, C.~P., {et~al.} 2013, \aj, 145, 10

\bibitem[{{de Vaucouleurs}(1948)}]{deV48}
{de Vaucouleurs}, G. 1948, Annales d'Astrophysique, 11, 247

\bibitem[{{Desmond} {et~al.}(2017){Desmond}, {Mao}, {Wechsler}, {Crain}, \&
  {Schaye}}]{Des++17}
{Desmond}, H., {Mao}, Y.-Y., {Wechsler}, R.~H., {Crain}, R.~A., \& {Schaye}, J.
  2017, \mnras, 471, L11

\bibitem[{{Eisenstein} {et~al.}(2011){Eisenstein}, {Weinberg}, {Agol},
  {Aihara}, {Allende Prieto}, {Anderson}, {Arns}, {Aubourg}, {Bailey},
  {Balbinot}, \& et~al.}]{Eis++11}
{Eisenstein}, D.~J., {Weinberg}, D.~H., {Agol}, E., {et~al.} 2011, \aj, 142, 72

\bibitem[{{Favole} {et~al.}(2018){Favole}, {Montero-Dorta}, {Prada},
  {Rodr{\'{\i}}guez-Torres}, \& {Schlegel}}]{Fav++18}
{Favole}, G., {Montero-Dorta}, A.~D., {Prada}, F., {Rodr{\'{\i}}guez-Torres},
  S.~A., \& {Schlegel}, D.~J. 2018, \mnras [\eprint[arXiv]{1802.01596}]

\bibitem[{{Foreman-Mackey} {et~al.}(2013){Foreman-Mackey}, {Hogg}, {Lang}, \&
  {Goodman}}]{For++13}
{Foreman-Mackey}, D., {Hogg}, D.~W., {Lang}, D., \& {Goodman}, J. 2013, \pasp,
  125, 306

\bibitem[{{Gallazzi} {et~al.}(2005){Gallazzi}, {Charlot}, {Brinchmann},
  {White}, \& {Tremonti}}]{Gal++05}
{Gallazzi}, A., {Charlot}, S., {Brinchmann}, J., {White}, S.~D.~M., \&
  {Tremonti}, C.~A. 2005, \mnras, 362, 41

\bibitem[{{Heymans} {et~al.}(2012){Heymans}, {Van Waerbeke}, {Miller}, {Erben},
  {Hildebrandt}, {Hoekstra}, {Kitching}, {Mellier}, {Simon}, {Bonnett},
  {Coupon}, {Fu}, {Harnois D{\'e}raps}, {Hudson}, {Kilbinger}, {Kuijken},
  {Rowe}, {Schrabback}, {Semboloni}, {van Uitert}, {Vafaei}, \&
  {Velander}}]{Hey++12}
{Heymans}, C., {Van Waerbeke}, L., {Miller}, L., {et~al.} 2012, \mnras, 427,
  146

\bibitem[{{Hirata} \& {Seljak}(2003)}]{H+S03}
{Hirata}, C. \& {Seljak}, U. 2003, \mnras, 343, 459

\bibitem[{{Hopkins} {et~al.}(2010{\natexlab{a}}){Hopkins}, {Bundy},
  {Hernquist}, {Wuyts}, \& {Cox}}]{Hop++10}
{Hopkins}, P.~F., {Bundy}, K., {Hernquist}, L., {Wuyts}, S., \& {Cox}, T.~J.
  2010{\natexlab{a}}, \mnras, 401, 1099

\bibitem[{{Hopkins} {et~al.}(2010{\natexlab{b}}){Hopkins}, {Croton}, {Bundy},
  {Khochfar}, {van den Bosch}, {Somerville}, {Wetzel}, {Keres}, {Hernquist},
  {Stewart}, {Younger}, {Genel}, \& {Ma}}]{Hop++10b}
{Hopkins}, P.~F., {Croton}, D., {Bundy}, K., {et~al.} 2010{\natexlab{b}}, \apj,
  724, 915

\bibitem[{{Huang} {et~al.}(2018{\natexlab{a}}){Huang}, {Leauthaud}, {Greene},
  {Bundy}, {Lin}, {Tanaka}, {Mandelbaum}, {Miyazaki}, \& {Komiyama}}]{Hua++18}
{Huang}, S., {Leauthaud}, A., {Greene}, J., {et~al.} 2018{\natexlab{a}},
  \mnras, 480, 521

\bibitem[{{Huang} {et~al.}(2018{\natexlab{b}}){Huang}, {Leauthaud}, {Hearin},
  {Behroozi}, {Bradshaw}, {Ardila}, {Speagle}, {Tenenti}, {Bundy}, {Greene},
  {Sifon}, \& {Bahcall}}]{Hua++18b}
{Huang}, S., {Leauthaud}, A., {Hearin}, A., {et~al.} 2018{\natexlab{b}}, ArXiv
  e-prints [\eprint[arXiv]{1811.01139}]

\bibitem[{{Huertas-Company} {et~al.}(2013{\natexlab{a}}){Huertas-Company},
  {Mei}, {Shankar}, {Delaye}, {Raichoor}, {Covone}, {Finoguenov}, {Kneib},
  {Le}, \& {Povic}}]{Hue++13}
{Huertas-Company}, M., {Mei}, S., {Shankar}, F., {et~al.} 2013{\natexlab{a}},
  \mnras, 428, 1715

\bibitem[{{Huertas-Company} {et~al.}(2013{\natexlab{b}}){Huertas-Company},
  {Shankar}, {Mei}, {Bernardi}, {Aguerri}, {Meert}, \& {Vikram}}]{Hue++13b}
{Huertas-Company}, M., {Shankar}, F., {Mei}, S., {et~al.} 2013{\natexlab{b}},
  \apj, 779, 29

\bibitem[{{Ivezic} {et~al.}(2008){Ivezic}, {Axelrod}, {Brandt}, {Burke},
  {Claver}, {Connolly}, {Cook}, {Gee}, {Gilmore}, {Jacoby}, {Jones}, {Kahn},
  {Kantor}, {Krabbendam}, {Lupton}, {Monet}, {Pinto}, {Saha}, {Schalk},
  {Schneider}, {Strauss}, {Stubbs}, {Sweeney}, {Szalay}, {Thaler}, {Tyson}, \&
  {LSST Collaboration}}]{Ive++08}
{Ivezic}, Z., {Axelrod}, T., {Brandt}, W.~N., {et~al.} 2008, Serbian
  Astronomical Journal, 176, 1

\bibitem[{{Juri{\'c}} {et~al.}(2015){Juri{\'c}}, {Kantor}, {Lim}, {Lupton},
  {Dubois-Felsmann}, {Jenness}, {Axelrod}, {Aleksi{\'c}}, {Allsman},
  {AlSayyad}, {Alt}, {Armstrong}, {Basney}, {Becker}, {Becla}, {Bickerton},
  {Biswas}, {Bosch}, {Boutigny}, {Carrasco Kind}, {Ciardi}, {Connolly},
  {Daniel}, {Daues}, {Economou}, {Chiang}, {Fausti}, {Fisher-Levine},
  {Freemon}, {Gee}, {Gris}, {Hernandez}, {Hoblitt}, {Ivezi{\'c}}, {Jammes},
  {Jevremovi{\'c}}, {Jones}, {Bryce Kalmbach}, {Kasliwal}, {Krughoff}, {Lang},
  {Lurie}, {Lust}, {Mullally}, {MacArthur}, {Melchior}, {Moeyens}, {Nidever},
  {Owen}, {Parejko}, {Peterson}, {Petravick}, {Pietrowicz}, {Price}, {Reiss},
  {Shaw}, {Sick}, {Slater}, {Strauss}, {Sullivan}, {Swinbank}, {Van Dyk},
  {Vuj{\v c}i{\'c}}, {Withers}, {Yoachim}, \& {LSST Project}}]{Jur++15}
{Juri{\'c}}, M., {Kantor}, J., {Lim}, K., {et~al.} 2015, ArXiv e-prints
  [\eprint[arXiv]{1512.07914}]

\bibitem[{{Kravtsov} {et~al.}(2018){Kravtsov}, {Vikhlinin}, \&
  {Meshcheryakov}}]{Kra++18}
{Kravtsov}, A.~V., {Vikhlinin}, A.~A., \& {Meshcheryakov}, A.~V. 2018,
  Astronomy Letters, 44, 8

\bibitem[{{Lani} {et~al.}(2013){Lani}, {Almaini}, {Hartley}, {Mortlock},
  {H{\"a}u{\ss}ler}, {Chuter}, {Simpson}, {van der Wel}, {Gr{\"u}tzbauch},
  {Conselice}, {Bradshaw}, {Cooper}, {Faber}, {Grogin}, {Kocevski},
  {Koekemoer}, \& {Lai}}]{Lan++13}
{Lani}, C., {Almaini}, O., {Hartley}, W.~G., {et~al.} 2013, \mnras, 435, 207

\bibitem[{{Leauthaud} {et~al.}(2016){Leauthaud}, {Bundy}, {Saito}, {Tinker},
  {Maraston}, {Tojeiro}, {Huang}, {Brownstein}, {Schneider}, \&
  {Thomas}}]{Lea++16}
{Leauthaud}, A., {Bundy}, K., {Saito}, S., {et~al.} 2016, \mnras, 457, 4021

\bibitem[{{Leauthaud} {et~al.}(2017){Leauthaud}, {Saito}, {Hilbert},
  {Barreira}, {More}, {White}, {Alam}, {Behroozi}, {Bundy}, {Coupon}, {Erben},
  {Heymans}, {Hildebrandt}, {Mandelbaum}, {Miller}, {Moraes}, {Pereira},
  {Rodr{\'{\i}}guez-Torres}, {Schmidt}, {Shan}, {Viel}, \&
  {Villaescusa-Navarro}}]{Lea++17}
{Leauthaud}, A., {Saito}, S., {Hilbert}, S., {et~al.} 2017, \mnras, 467, 3024

\bibitem[{{Macci{\`o}} {et~al.}(2008){Macci{\`o}}, {Dutton}, \& {van den
  Bosch}}]{Mac++08}
{Macci{\`o}}, A.~V., {Dutton}, A.~A., \& {van den Bosch}, F.~C. 2008, \mnras,
  391, 1940

\bibitem[{{Mandelbaum} {et~al.}(2018){Mandelbaum}, {Miyatake}, {Hamana},
  {Oguri}, {Simet}, {Armstrong}, {Bosch}, {Murata}, {Lanusse}, {Leauthaud},
  {Coupon}, {More}, {Takada}, {Miyazaki}, {Speagle}, {Shirasaki}, {Sif{\'o}n},
  {Huang}, {Nishizawa}, {Medezinski}, {Okura}, {Okabe}, {Czakon}, {Takahashi},
  {Coulton}, {Hikage}, {Komiyama}, {Lupton}, {Strauss}, {Tanaka}, \&
  {Utsumi}}]{Man++18}
{Mandelbaum}, R., {Miyatake}, H., {Hamana}, T., {et~al.} 2018, \pasj, 70, S25

\bibitem[{{Mandelbaum} {et~al.}(2016){Mandelbaum}, {Wang}, {Zu}, {White},
  {Henriques}, \& {More}}]{Man++16}
{Mandelbaum}, R., {Wang}, W., {Zu}, Y., {et~al.} 2016, \mnras, 457, 3200

\bibitem[{{Maraston} {et~al.}(2013){Maraston}, {Pforr}, {Henriques}, {Thomas},
  {Wake}, {Brownstein}, {Capozzi}, {Tinker}, {Bundy}, {Skibba}, {Beifiori},
  {Nichol}, {Edmondson}, {Schneider}, {Chen}, {Masters}, {Steele}, {Bolton},
  {York}, {Weaver}, {Higgs}, {Bizyaev}, {Brewington}, {Malanushenko},
  {Malanushenko}, {Snedden}, {Oravetz}, {Pan}, {Shelden}, \&
  {Simmons}}]{Mar++13}
{Maraston}, C., {Pforr}, J., {Henriques}, B.~M., {et~al.} 2013, \mnras, 435,
  2764

\bibitem[{{Marshall} {et~al.}(2016){Marshall}, {Verma}, {More}, {Davis},
  {More}, {Kapadia}, {Parrish}, {Snyder}, {Wilcox}, {Baeten}, {Macmillan},
  {Cornen}, {Baumer}, {Simpson}, {Lintott}, {Miller}, {Paget}, {Simpson},
  {Smith}, {K{\"u}ng}, {Saha}, \& {Collett}}]{Mar++16}
{Marshall}, P.~J., {Verma}, A., {More}, A., {et~al.} 2016, \mnras, 455, 1171

\bibitem[{{McCavana} {et~al.}(2012){McCavana}, {Micic}, {Lewis}, {Sinha},
  {Sharma}, {Holley-Bockelmann}, \& {Bland-Hawthorn}}]{McC++12}
{McCavana}, T., {Micic}, M., {Lewis}, G.~F., {et~al.} 2012, \mnras, 424, 361

\bibitem[{{Montero-Dorta} {et~al.}(2016){Montero-Dorta}, {Bolton},
  {Brownstein}, {Swanson}, {Dawson}, {Prada}, {Eisenstein}, {Maraston},
  {Thomas}, {Comparat}, {Chuang}, {McBride}, {Favole}, {Guo},
  {Rodr{\'{\i}}guez-Torres}, \& {Schneider}}]{Mon++16}
{Montero-Dorta}, A.~D., {Bolton}, A.~S., {Brownstein}, J.~R., {et~al.} 2016,
  \mnras, 461, 1131

\bibitem[{{More} {et~al.}(2015){More}, {Miyatake}, {Mandelbaum}, {Takada},
  {Spergel}, {Brownstein}, \& {Schneider}}]{Mor++15}
{More}, S., {Miyatake}, H., {Mandelbaum}, R., {et~al.} 2015, \apj, 806, 2

\bibitem[{{Naab} {et~al.}(2009){Naab}, {Johansson}, \& {Ostriker}}]{Naa++09}
{Naab}, T., {Johansson}, P.~H., \& {Ostriker}, J.~P. 2009, \apjl, 699, L178

\bibitem[{{Navarro} {et~al.}(1997){Navarro}, {Frenk}, \& {White}}]{NFW97}
{Navarro}, J.~F., {Frenk}, C.~S., \& {White}, S.~D.~M. 1997, \apj, 490, 493

\bibitem[{{Newman} {et~al.}(2014){Newman}, {Ellis}, {Andreon}, {Treu},
  {Raichoor}, \& {Trinchieri}}]{New++14}
{Newman}, A.~B., {Ellis}, R.~S., {Andreon}, S., {et~al.} 2014, \apj, 788, 51

\bibitem[{{Newman} {et~al.}(2012){Newman}, {Ellis}, {Bundy}, \&
  {Treu}}]{New++12}
{Newman}, A.~B., {Ellis}, R.~S., {Bundy}, K., \& {Treu}, T. 2012, \apj, 746,
  162

\bibitem[{{Nipoti} {et~al.}(2012){Nipoti}, {Treu}, {Leauthaud}, {Bundy},
  {Newman}, \& {Auger}}]{Nip++12}
{Nipoti}, C., {Treu}, T., {Leauthaud}, A., {et~al.} 2012, \mnras, 422, 1714

\bibitem[{{Oguri} {et~al.}(2018){Oguri}, {Lin}, {Lin}, {Nishizawa}, {More},
  {More}, {Hsieh}, {Medezinski}, {Miyatake}, {Jian}, {Lin}, {Takada}, {Okabe},
  {Speagle}, {Coupon}, {Leauthaud}, {Lupton}, {Miyazaki}, {Price}, {Tanaka},
  {Chiu}, {Komiyama}, {Okura}, {Tanaka}, \& {Usuda}}]{Ogu++18}
{Oguri}, M., {Lin}, Y.-T., {Lin}, S.-C., {et~al.} 2018, \pasj, 70, S20

\bibitem[{{Papovich} {et~al.}(2012){Papovich}, {Bassett}, {Lotz}, {van der
  Wel}, {Tran}, {Finkelstein}, {Bell}, {Conselice}, {Dekel}, {Dunlop}, {Guo},
  {Faber}, {Farrah}, {Ferguson}, {Finkelstein}, {H{\"a}ussler}, {Kocevski},
  {Koekemoer}, {Koo}, {McGrath}, {McLure}, {McIntosh}, {Momcheva}, {Newman},
  {Rudnick}, {Weiner}, {Willmer}, \& {Wuyts}}]{Pap++12}
{Papovich}, C., {Bassett}, R., {Lotz}, J.~M., {et~al.} 2012, \apj, 750, 93

\bibitem[{{Reid} {et~al.}(2016){Reid}, {Ho}, {Padmanabhan}, {Percival},
  {Tinker}, {Tojeiro}, {White}, {Eisenstein}, {Maraston}, {Ross},
  {S{\'a}nchez}, {Schlegel}, {Sheldon}, {Strauss}, {Thomas}, {Wake}, {Beutler},
  {Bizyaev}, {Bolton}, {Brownstein}, {Chuang}, {Dawson}, {Harding}, {Kitaura},
  {Leauthaud}, {Masters}, {McBride}, {More}, {Olmstead}, {Oravetz}, {Nuza},
  {Pan}, {Parejko}, {Pforr}, {Prada}, {Rodr{\'{\i}}guez-Torres},
  {Salazar-Albornoz}, {Samushia}, {Schneider}, {Sc{\'o}ccola}, {Simmons}, \&
  {Vargas-Magana}}]{Rei++16}
{Reid}, B., {Ho}, S., {Padmanabhan}, N., {et~al.} 2016, \mnras, 455, 1553

\bibitem[{{Saracco} {et~al.}(2017){Saracco}, {Gargiulo}, {Ciocca}, \&
  {Marchesini}}]{Sar++17}
{Saracco}, P., {Gargiulo}, A., {Ciocca}, F., \& {Marchesini}, D. 2017, \aap,
  597, A122

\bibitem[{{Schaye} {et~al.}(2015){Schaye}, {Crain}, {Bower}, {Furlong},
  {Schaller}, {Theuns}, {Dalla Vecchia}, {Frenk}, {McCarthy}, {Helly},
  {Jenkins}, {Rosas-Guevara}, {White}, {Baes}, {Booth}, {Camps}, {Navarro},
  {Qu}, {Rahmati}, {Sawala}, {Thomas}, \& {Trayford}}]{Sch++15}
{Schaye}, J., {Crain}, R.~A., {Bower}, R.~G., {et~al.} 2015, \mnras, 446, 521

\bibitem[{{Schlegel} {et~al.}(2009){Schlegel}, {White}, \&
  {Eisenstein}}]{SWE09}
{Schlegel}, D., {White}, M., \& {Eisenstein}, D. 2009, in ArXiv Astrophysics
  e-prints, Vol. 2010, astro2010: The Astronomy and Astrophysics Decadal Survey

\bibitem[{{Sersic}(1968)}]{Ser68}
{Sersic}, J.~L. 1968, {Atlas de galaxias australes} (Cordoba, Argentina:
  Observatorio Astronomico)

\bibitem[{{Shankar} {et~al.}(2015){Shankar}, {Buchan}, {Rettura}, {Bouillot},
  {Moreno}, {Licitra}, {Bernardi}, {Huertas-Company}, {Mei}, {Ascaso}, {Sheth},
  {Delaye}, \& {Raichoor}}]{Sha++15}
{Shankar}, F., {Buchan}, S., {Rettura}, A., {et~al.} 2015, \apj, 802, 73

\bibitem[{{Shankar} {et~al.}(2014{\natexlab{a}}){Shankar}, {Guo}, {Bouillot},
  {Rettura}, {Meert}, {Buchan}, {Kravtsov}, {Bernardi}, {Sheth}, {Vikram},
  {Marchesini}, {Behroozi}, {Zheng}, {Maraston}, {Ascaso}, {Lemaux}, {Capozzi},
  {Huertas-Company}, {Gal}, {Lubin}, {Conselice}, {Carollo}, \&
  {Cattaneo}}]{Sha++14b}
{Shankar}, F., {Guo}, H., {Bouillot}, V., {et~al.} 2014{\natexlab{a}}, \apjl,
  797, L27

\bibitem[{{Shankar} {et~al.}(2014{\natexlab{b}}){Shankar}, {Mei},
  {Huertas-Company}, {Moreno}, {Fontanot}, {Monaco}, {Bernardi}, {Cattaneo},
  {Sheth}, {Licitra}, {Delaye}, \& {Raichoor}}]{Sha++14}
{Shankar}, F., {Mei}, S., {Huertas-Company}, M., {et~al.} 2014{\natexlab{b}},
  \mnras, 439, 3189

\bibitem[{{Shankar} {et~al.}(2018){Shankar}, {Sonnenfeld}, {Grylls}, {Zanisi},
  {Nipoti}, {Chae}, {Bernardi}, {Petrillo}, {Huertas-Company}, {Mamon}, \&
  {Buchan}}]{Sha++18}
{Shankar}, F., {Sonnenfeld}, A., {Grylls}, P., {et~al.} 2018, \mnras, 475, 2878

\bibitem[{{Sonnenfeld} \& {Leauthaud}(2018)}]{S+L18}
{Sonnenfeld}, A. \& {Leauthaud}, A. 2018, \mnras, 477, 5460

\bibitem[{{Sonnenfeld} {et~al.}(2018){Sonnenfeld}, {Leauthaud}, {Auger},
  {Gavazzi}, {Treu}, {More}, \& {Komiyama}}]{Son++18}
{Sonnenfeld}, A., {Leauthaud}, A., {Auger}, M.~W., {et~al.} 2018, \mnras, 481,
  164

\bibitem[{{Tanaka}(2015)}]{Tan15}
{Tanaka}, M. 2015, \apj, 801, 20

\bibitem[{{Tanaka} {et~al.}(2018){Tanaka}, {Coupon}, {Hsieh}, {Mineo},
  {Nishizawa}, {Speagle}, {Furusawa}, {Miyazaki}, \& {Murayama}}]{Tan++18}
{Tanaka}, M., {Coupon}, J., {Hsieh}, B.-C., {et~al.} 2018, \pasj, 70, S9

\bibitem[{{Tinker} {et~al.}(2017){Tinker}, {Brownstein}, {Guo}, {Leauthaud},
  {Maraston}, {Masters}, {Montero-Dorta}, {Thomas}, {Tojeiro}, {Weiner},
  {Zehavi}, \& {Olmstead}}]{Tin++17}
{Tinker}, J.~L., {Brownstein}, J.~R., {Guo}, H., {et~al.} 2017, \apj, 839, 121

\bibitem[{{Trujillo} {et~al.}(2006){Trujillo}, {F{\"o}rster Schreiber},
  {Rudnick}, {Barden}, {Franx}, {Rix}, {Caldwell}, {McIntosh}, {Toft},
  {H{\"a}ussler}, {Zirm}, {van Dokkum}, {Labb{\'e}}, {Moorwood},
  {R{\"o}ttgering}, {van der Wel}, {van der Werf}, \& {van
  Starkenburg}}]{Tru++06}
{Trujillo}, I., {F{\"o}rster Schreiber}, N.~M., {Rudnick}, G., {et~al.} 2006,
  \apj, 650, 18

\bibitem[{{van der Wel} {et~al.}(2008){van der Wel}, {Holden}, {Zirm}, {Franx},
  {Rettura}, {Illingworth}, \& {Ford}}]{vdW++08}
{van der Wel}, A., {Holden}, B.~P., {Zirm}, A.~W., {et~al.} 2008, \apj, 688, 48

\bibitem[{{van Dokkum} {et~al.}(2008){van Dokkum}, {Franx}, {Kriek}, {Holden},
  {Illingworth}, {Magee}, {Bouwens}, {Marchesini}, {Quadri}, {Rudnick},
  {Taylor}, \& {Toft}}]{van++08}
{van Dokkum}, P.~G., {Franx}, M., {Kriek}, M., {et~al.} 2008, \apjl, 677, L5

\bibitem[{{Vogelsberger} {et~al.}(2014){Vogelsberger}, {Genel}, {Springel},
  {Torrey}, {Sijacki}, {Xu}, {Snyder}, {Nelson}, \& {Hernquist}}]{Vog++14}
{Vogelsberger}, M., {Genel}, S., {Springel}, V., {et~al.} 2014, \mnras, 444,
  1518

\bibitem[{{Westera} {et~al.}(2002){Westera}, {Lejeune}, {Buser}, {Cuisinier},
  \& {Bruzual}}]{Wes++02}
{Westera}, P., {Lejeune}, T., {Buser}, R., {Cuisinier}, F., \& {Bruzual}, G.
  2002, \aap, 381, 524

\bibitem[{{Yoon} {et~al.}(2017){Yoon}, {Im}, \& {Kim}}]{Yoo++17}
{Yoon}, Y., {Im}, M., \& {Kim}, J.-W. 2017, \apj, 834, 73

\end{thebibliography}

\end{document}